\documentclass[11pt,a4paper]{article}
\usepackage[T1]{fontenc}
\pdfoutput=1
\usepackage{jheppub}
\usepackage{amsthm}
\usepackage{bbm}

\usepackage{hyperref}

\usepackage{physics}
\usepackage{xcolor}
\usepackage{mathtools}
\usepackage{tensor}
\usepackage{empheq}

\usepackage{tikz}
\usepackage{graphicx}
\usepackage{subfigure}
\usetikzlibrary{calc,fadings,decorations.pathreplacing}
\usepackage{verbatim}
\usepackage{tikz-3dplot}
\usetikzlibrary{arrows,calc,backgrounds,patterns}

\usepackage{url}

\DeclareMathOperator{\MyProd}{\scalebox{1.4}{$\mathrm{I\kern-0.2ex I}$}}
\newcommand\fft[2]{\frac{#1}{#2}}

\def\i{\mathrm{i}}

\pgfmathsetmacro{\r}{1.6} 
\pgfmathsetmacro{\h}{0.9*\r} 
\pgfmathsetmacro{\rlat}{1.2*sqrt(\r*\r-\h*\h)}
\pgfmathsetmacro{\gap}{1.5} 

\tdplotsetmaincoords{-20}{0}

\tikzset{%
  >=latex, 
  inner sep=0pt,%
  outer sep=2pt,%
  mark coordinate/.style={inner sep=0pt,outer sep=0pt,minimum size=3pt,
    fill=black,circle}%
}

\title{ The giant graviton expansion in AdS$_5\times$SE$_5$  }

\author[a]{\small{Alfredo Gonz\'alez Lezcano}}
\author[b]{ \small{Leopoldo A. Pando Zayas}}
\author[c]{\small{and Augniva Ray}}

\emailAdd{alfredo.gonzalez@unipd.it
, lpandoz@umich.edu, augniva@gmail.com}

\affiliation[a]{Dipartimento di Fisica e Astronomia ``Galileo Galilei'', Universit\`a di Padova and INFN Sezione di Padova, Via Marzolo 8, 35131 Padova, Italy. }

\affiliation[b]
{Leinweber Institute for Theoretical Physics, 
University of Michigan, Ann Arbor, MI 48109, USA}

\affiliation[c]{Department of Physics and Astronomy \& Center for Theoretical Physics, Seoul National University, Seoul 08826, Korea}

\abstract{
We study giant graviton-like D3-branes as probe configurations in type IIB supergravity backgrounds AdS$_5\times$SE$_5$, with emphasis on the structure of the five-dimensional Sasaki-Einstein manifolds SE$_5$. These configurations wrap supersymmetric three-cycles in SE$_5$ and rotate at the speed of light along the Reeb direction. We formulate the general problem in terms of the transverse K\"ahler potential and show that configurations carrying maximal angular momentum can be described by loci where the transverse K\"ahler potential diverges in suitably chosen coordinates. We quantize a particular set of excitations, similar to those considered in the AdS$_5\times S^5$ case, and show that they are governed by a Fock-Darwin problem with a conical deficit, which generalizes the Landau problem obtained for $S^5$. Succinctly, the information about the geometry of SE$_5$ is encoded in the form of the external potential and the conical deficit. 
We compute a protected contribution to the superconformal index arising from quantized fluctuations of supersymmetric giant graviton configurations.
For the explicit case of AdS$_5\times T^{1,1}$, we recover the finite-$N$ protected index of the dual quiver ${\cal N}=1$ superconformal field theory in the sector captured by a non-Abelian generalization of the quantized degrees of freedom. }


\arxivnumber{}

\begin{document}
\today 
\maketitle


\section{Introduction}

One of the most insightful aspects of the AdS/CFT correspondence is that it provides whole classes of dualities. One particularly powerful instance is the duality between type IIB supergravity backgrounds on AdS$_5\times$ SE$_5$, where SE$_5$ is a Sasaki-Einstein manifold,  with $N$ units of five-form flux and certain quiver gauge theories with gauge group a product of $SU(N)$ factors and ${\cal N}=1$ superconformal symmetry \cite{Benvenuti:2004dy}. The prototypical example of the AdS/CFT correspondence, the duality between IIB strings in AdS$_5\times S^5$ and ${\cal N}=4$ supersymmetric Yang-Mills is but one instance of this large class of dualities \cite{Maldacena:1997re}. 
We can distinguish between what are properties of AdS$_5\times S^5$ and what are generic properties of gravitational degrees of freedom.

In the context of a quantum theory,  partition functions and their variations play a central role in understanding the degrees of freedom. The  reinterpretations of some of these counting functions in terms of expansions dubbed  giant graviton expansions \cite{Arai:2019aou, Arai:2020qaj,Imamura:2021ytr,Gaiotto:2021xce} (see also some precedence-setting work \cite{Bourdier:2015wda,Drukker:2015spa} and \cite{Choi:2022ovw, Kim:2024ucf, Chen:2024erz} for recent progress on the relation between giant gravitons and black hole entropy) has generated considerable interest. The generic form of such expansion (considering only one fugacity $q$ for simplicity) is 

\begin{equation}
	\label{Eq:GGExpansions}
	{\cal I}_N(q)= {\cal I}_\infty \left( \sum\limits_{m=0}^\infty q^{m\, N} \mathcal{J}_m(q)\right),
\end{equation}
where the so-called giant graviton partition $\mathcal{J}_m(q)$ is independent of the rank of the gauge group, $N$.

Given that on the gravitational side, the effective tension of D3 branes wrapping certain 3-cycles in the supergravity solution is precisely $N$, a natural holographic interpretation in terms of giant gravitons is anticipated. This intuition has been proven correct and shows that the relevant gravitational degrees of freedom are giant gravitons. 

Identifying giant gravitons as central  gravitational degrees of freedom opens a new conceptual stage. In the context of the AdS/CFT correspondence, various enumerating problems are reformulated on the field theory side where the enumeration is more direct. This indirect route leaves unanswered the question of identifying the degrees of freedom on the gravitational side of the correspondence. The cleanest example of such outsourcing is the black hole entropy microstate counting. In this manuscript we isolate universal gravitational building blocks underlying protected sectors in a large class of AdS/CFT dual pairs, focusing on their realization in terms of supersymmetric D3 branes. Throughout the paper, we consider calibrated D3-branes wrapping supersymmetric three-cycles associated with toric holomorphic divisors of the Calabi-Yau cone. These configurations preserve the supercharge used to define the index (equivalently, one quarter of the supersymmetry of the AdS$_5\times$SE$_5$ background) and provide the gravitational sector that we quantize. Our terminology is therefore adapted to the ${\cal N}=1$ Sasaki-Einstein setting and is not intended to mirror the finer BPS hierarchy familiar from the maximally supersymmetric AdS$_5\times S^5$ case.
There have been a number of recent works exploring the gravitational side of giant graviton expansions of partition functions. Most prominently are discussions in terms of probe branes  \cite{Lee:2023iil,Eleftheriou:2023jxr,Beccaria:2024vfx,Gautason:2024nru,Lee:2024hef,Eleftheriou:2025lac}. Some of these works, which we briefly review, have succeeded in reproducing the corresponding finite-$N$ protected index. An interpretation in terms of fully back-reacted geometries leading likewise to a finite-$N$ reproduction of the index was also given in \cite{Deddo:2024liu}. One of our goals in this paper is to investigate the extent to which these results extend to the broader class of AdS$_5\times$SE$_5/{\cal N}=1$ SCFT dual pairs. Connections between the superconformal index and the gravitational dual background were first discussed in \cite{Kinney:2005ej} for ${\cal N}=4$ SYM. In particular, the term ${\cal I}_\infty$ in the index was identified through a Kaluza-Klein analysis of the supergravity spectrum; extensions to ${\cal N}=1$ backgrounds were subsequently developed, for example, in \cite{Gadde:2010en,Nakayama:2006ur,Eager:2012hx}.

In light of these developments, we explore generic structures that allow us to quantize local fluctuations about giant graviton-like configurations. To this end, we first revisit the giant graviton on $S^5$ to emphasize the generic Sasaki-Einstein structure. Centering the analysis on the complex structure of the cone over the SE$_5$ as pioneered by Mikhailov in \cite{Mikhailov:2000ya} (see also \cite{Beasley:2002xv}) plays a central role in establishing the classical configuration. This point of view allowed us to retrieve known results for the case of giant gravitons on AdS$_5\times S^5$ explicitly using the structure of $S^5$ as a $U(1)$ fiber over $\mathbb{CP}^2$. The distinctive role played by the maximal giant configurations has been established via localization arguments \cite{Eleftheriou:2023jxr,Eleftheriou:2025lac}, we therefore focus on  analogous configurations and quantize certain fluctuations around them. A key point of view adopted in this work is that the relevant quantum mechanics governing these configurations localizes to neighborhoods of toric fixed points of the Calabi–Yau cone, leading to effective Landau-type systems.

In the generic case of AdS$_5\times$SE$_5$, we leverage the role of the complex structure in determining giant graviton configurations, which we connect to the transverse K\"ahler-Einstein potential. We also establish that the dynamics of certain local fluctuations about maximal giant graviton-like configurations can be effectively described by Landau-type problems, and generalizations thereof. In particular, we model the Hilbert space of such fluctuations associated with multi-giant gravitons using matrix model descriptions familiar from studies of the Quantum Hall effect. The example that we work out in more detail is the case of SE$_5 = T^{1,1}$, where we recover the corresponding finite-$N$ protected index in this setting, thereby illustrating the mechanism in a controlled example. We use this explicit case to extract universal features of the quantization procedure.

The rest of the manuscript is organized as follows. Section \ref{Sec:ReviewS5} revisits the prototypical giant graviton on AdS$_5\times S^5$ in a language that emphasizes the structure of $S^5$ as a SE$_5$ manifold. In section \ref{sec:GeneralKE} we present the general analysis for giant gravitons in AdS$_5\times$SE$_5$. In section \ref{sec:ExampleT11}, we study in detail the giant graviton index at finite $N$ for the case of AdS$_5\times T^{1,1}$ dual to the Klebanov-Witten theory, focusing on the corresponding protected sector. We conclude in section \ref{Sec:Conclusion}. We provide some relevant background material on the Quantum Hall effect in Appendix \ref{App:QHE}.

\section{Revisiting the giant graviton in AdS$_5\times S^5$}\label{Sec:ReviewS5}

The type IIB supergravity background with metric AdS$_5\times S^5$ and supported by $N$ units of five-form flux admits classical configurations of wrapped branes known as giant gravitons. The original giant graviton classical configuration \cite{McGreevy:2000cw}  describes a  D3 brane wrapping an $S^3$ and rotating along the Reeb direction of an $S^5$. The solution is further supported by the Wess-Zumino term.   This solution also extends along the temporal direction in AdS$_5$ and carries angular momentum that is bounded by the size of $S^5$.

The action of the classical configuration with maximum angular momentum is  proportional to $N$ due to a precise combination of the D3 brane tension and  other constants of the solution. This proportionality becomes a  key ingredient in interpreting field-theoretic expansion of supersymmetric partition functions involving powers of $q^N$ which are the so-called giant graviton expansions discussed in \cite{Bourdier:2015wda,Arai:2020qaj,Imamura:2021ytr,Gaiotto:2021xce} and schematically displayed in \eqref{Eq:GGExpansions}. The study of fluctuations around these classical configurations was initiated in \cite{Das:2000st} (see also \cite{Arapoglu:2003ti,Ouyang:2002vg}). More recently,  the case of D3 branes has been revisited in \cite{Beccaria:2024vfx, Gautason:2024nru} and  a complete description of all the fluctuations for the D3, M2 and M5 brane giant gravitons has been provided in \cite{Deddo:2025lfm}.

In this manuscript we explore the giant graviton expansions from the holographic point, of view in various new directions. As a natural starting point we first revisit various aspects of the canonical holographic giant graviton expansion in AdS$_5\times S^5$ with some emphasis on extensions to SE$_5$.

\subsection{ Reformulating the problem for the sphere giant graviton} \label{subsec:S5GGreformulated}

Let us reformulate the problem of the giant graviton and its fluctuations  in a language more suitable for generalizations to the SE$_5$ case. We cast  $S^5$ as a $U(1)$ fibration over $\mathbb{CP}^2$.
We consider the AdS$_5\times S^5$ metric in the form
\begin{align}
	d s^2 & = L^2d s^2_{\text{AdS}_5} + L^2(d \Sigma_2^2 + \sigma^2 )\,,\\
	d s^2_{\text{AdS}_5} & = -\left(1+\frac{r^2}{L^2}\right)dt^2 + \frac{dr^2}{1+\frac{r^2}{L^2}} + r^2 d\hat{\Omega}_3^2 \, , \label{Eq:AdS5} \\ \label{eq:dS2S5}
	d \Sigma_2^2 +\sigma^2& = d \xi^2 + \frac{1}{4} \sin^2 \xi \left(s_1^2 + s_2^2\right) + \frac{1}{4} \sin^2 \xi \cos^2 \xi s_3^2 +\left(d \tau + \frac{1}{2} \sin^2 \xi s_3\right)^2  \, ,
\end{align}
where $\sigma =d\tau +\frac{1}{2}\sin^2\xi s_3$.
Here $s_{1,2,3}$  are a set of left-invariant 1-forms on $SU(2)$:
\begin{align}
	s_1 & = \sin \theta_3 d \theta_1 - \cos \theta_3 \sin \theta_1 d \theta_2\,, \\
	s_2 & = \cos \theta_3 d \theta_1 + \sin \theta_3 \sin 
	\theta_1 d \theta_2\,, \\
	s_3 & = d \theta_3 + \cos \theta_1 d \theta_2 \, ,
\end{align}
where $0 \leq \theta_1 \leq \pi, \, \, 0 \leq \theta_2 \leq 2 \pi$ and  $0 \leq \theta_3 \leq 4 \pi$ and $ds_i  + \epsilon_{ijk} s_j \wedge s_k = 0$.
Let us first discuss the choice of coordinates with a focus on the complex structure. The coordinates on $\mathbb{CP}^2$ are $(\xi, \theta_1,\theta_2, \theta_3)$. Note that $(\theta_1,\theta_2, \theta_3)$ effectively parametrize a 3-sphere in the Hopf parametrization with fiber $\theta_3$ and 2-sphere given by $(\theta_1, \theta_2)$. 
We would like to understand the giant graviton in terms of the transverse K\"ahler structure of $S^5$.
Consider now the metric for $\mathbb{CP}^2$ obtained from the metric induced on the unit  $S^5$ viewed as embedded in $\mathbb{C}^3$. We have:
\begin{align}
	L^2 &  =  |Z_1|^2+ |Z_2|^2+ |Z_3|^2,  \\ \label{eq:dS5}
	ds^2_{S^5}  & = |dZ_1|^2+ |dZ_2|^2+ |dZ_3|^2 \,.
\end{align}
The complex coordinates of $\mathbb{C}^3$ in terms of the angular variables used in \eqref{eq:dS2S5} are:
\begin{align} \label{eq:ZAdef}
	\begin{split}
		Z_1 &= L \sin\xi
		\cos\frac{\theta_1}{2}
		\, e^{i\left(\tau+\frac{\theta_2+\theta_3}{2}\right)}, \\
		Z_2 &= L \sin\xi
		\sin\frac{\theta_1}{2}
		\, e^{i\left(\tau+\frac{\theta_2-\theta_3}{2}\right)}, \\
		Z_3 &= L \cos\xi\, e^{i\tau}.
	\end{split}
\end{align}
Note that the action
\begin{align}
	Z_A \rightarrow {\rm e}^{i \frac{t}{L}} Z_{A}, \, \quad A=1,2,3\,,
\end{align} where $t$ is the AdS$_5$ time coordinate, maps $S^5$ into itself and defines a one-parameter family of transformations that determines a $U(1)$ isometry. 
By projecting the metric \eqref{eq:dS5} of $S^5$ orthogonally onto the orbits of this $U(1)$ isometry, we can obtain the metric \eqref{eq:KT} below for $\mathbb{CP}^2$. Let us see this more explicitly.
Defining the homogeneous complex coordinates:
\begin{align}
	\zeta^1 & = \frac{Z_1}{Z_3}=\tan \xi \cos \frac{\theta_1}{2} {\rm e}^{i \frac{\theta_3 +\theta_2}{2}} \, , \\
	\zeta^2 & = \frac{Z_2}{Z_3}=\tan \xi \sin \frac{\theta_1}{2}{\rm e}^{i \frac{\theta_2- \theta_3}{2}} \, ,
\end{align}
we can then write:
\begin{align} \label{eq:KT}
	d \Sigma_2^2  & = \frac{\partial^2 K(\zeta^a, \, \bar{\zeta}^{\bar{b}})}{\partial \zeta^a  \partial \bar{\zeta}^{\bar{b}}} d \zeta^a d \bar{\zeta}^{\bar{b}} \, , \quad a, \, b = 1,2 \, ,
	\\
	K(\zeta^a, \, \bar{\zeta}^{\bar{b}}) & = L^2 \log \left(1+ L^{-2}(|\zeta^1|^2 + |\zeta^2|^2)\right)\, ,
\end{align} which manifestly exhibits the K\"ahler structure of $\mathbb{CP}^2$. We have used the subindex $T$ to emphasize the fact that here $\mathbb{CP}^2$ appears as the transverse K\"ahler manifold. In fact the transverse K\"ahler form is given by:
\begin{align} \label{eq:JT}
	J_T & = i \partial \bar{\partial} K\, .
\end{align}
Therefore we can see $S^5$ as a $U(1)$ fibration over a K\"ahler manifold with connection given by $\sigma = d \tau + \frac{1}{2} \sin^2 \xi s_3$, such that $d\sigma =2 J_{\text{T}}$. 
This is to say that $S^5$ is a Sasaki-Einstein manifold whose metric can be written locally as:
\begin{align}
	d s^2_{\text{SE}} & = d\Sigma_2^2 + \left(d \tau + a_T\right)^2 \,, \\
	a_T & = \frac{1}{2} \sin^2 \xi \, s_3 \ .   
\end{align}
The Reeb vector field is $\frac{\partial}{\partial \tau}$. The corresponding metric cone has line element:
\begin{align}
	d s^2_{C} & = d \rho^2 + \rho^2 d s_{\text{SE}_5}^2\, ,
\end{align}
which is K\"ahler. Mikhailov's embedding defines giant gravitons as \cite{Mikhailov:2000ya}:
\begin{eqnarray} \label{Eq:MikhailovEmb}
	|Z_1|^2+ |Z_2|^2+ |Z_3|^2 & =&L^2, \, \nonumber \\
	F({\rm e}^{- i \frac{t}{L}}Z_1, {\rm e}^{- i \frac{t}{L}}Z_2, {\rm e}^{- i \frac{t}{L}}Z_3) & = &0\,,
\end{eqnarray}
where $F(...)$ is a holomorphic function.  This point of view of giants as holomorphic hypersurfaces intersected with the base of the cone, was further developed in the language of moduli space in \cite{Beasley:2002xv}, including a discussion of giants in $T^{1,1}$. Other insights in the context of $S^5$  and particularly relevant for  $\frac{1}{8}$-BPS configurations were presented in \cite{Biswas:2006tj}.

The framework above \eqref{Eq:MikhailovEmb} suggests that the natural rotating coordinate required to define the embedding of a giant graviton should be the one whose direction is given by the Reeb vector. To simplify the analysis, let us choose $F(Z_1, Z_2, Z_3) = Z_3- b$ with $b \in \mathbb{R}$, we have:
\begin{align} \label{eq:Z3b}
	Z_3 & = b \, {\rm e}^{i \frac{t}{L}}\,, \quad  \textrm{with} \quad b=L \cos \xi\,,
\end{align} consistent with \eqref{Eq:MikhailovEmb}.
We choose the following embedding functions:
\begin{align}
	\tau & = \tau(t), \, \quad \xi = \xi(t), \quad 
	\sigma^{a}  = \text{arg}(\zeta^a)\, , \quad \sigma^3 = \theta_1\,, 
\end{align}
which gives us the following induced metric
\begin{align}
	g_{\text{D}3} & =\left(
	\begin{array}{cccc}
		\dot{\xi}^2-\frac{r^2}{L^2}+\frac{\dot{r}^2}{\frac{r^2}{L^2}+1}+\dot{\tau}^2-1 & 0 & \frac{\dot{\tau}}{2} \cos \theta_1 \sin ^2\xi   & \frac{\dot{\tau}}{2} \sin ^2\xi   \\
		0 & \frac{1}{4} \sin ^2\xi  & 0 & 0 \\
		\frac{\dot{\tau}}{2} \cos \theta_1 \sin ^2\xi  & 0 & \frac{1}{4} \sin ^2\xi & \frac{1}{4} \cos \theta_1 \sin ^2\xi  \\
		\frac{\dot{\tau}}{2} \sin ^2\xi  & 0 & \frac{1}{4} \cos \theta_1 \sin ^2\xi  & \frac{1}{4} \sin ^2\xi  \\
	\end{array}
	\right)\, ,
\end{align} and 
the corresponding DBI action is 
\begin{align}
	S_{\text{DBI}} &  = -2 \pi ^2 T_3 L^3\sin ^3\xi  \sqrt{1+\frac{r^2}{L^2}-\dot{\xi}^2-\dot{\tau}^2\cos ^2\xi -\frac{\dot{r}^2}{\frac{r^2}{L^2}+1}}\, .
\end{align}
We consider the following choice of 4-form potential
\begin{align}
	\begin{split}
		F_5&={\rm Vol}(S^5), \\
		&=\frac{L^4}{8} \sin \theta_1 d\left(\sin ^4\xi \right)\wedge  d\theta_1\wedge d\theta_2\wedge d\theta_3 \wedge d\tau\,,  \\
		C_4&= \frac{L^4}{8} \sin \theta_1 \sin ^4\xi  d\theta_1\wedge d\theta_2\wedge d\theta_3\wedge d\tau + d[\Lambda_{(3)}] \,,\\
		\Lambda_{(3)} & = \frac{L^4}{8}\cos \theta_1 d \theta_2 \wedge d \theta_3 \wedge d \tau \, ,
	\end{split}
\end{align}
where we have chosen a gauge for the 4-form potential such that there will be no singular term near the maximal giant configuration.
The WZ term is then given by
\begin{align} 
	S_{\text{WZ}} & = 2 \pi ^2 T_3 L^4\int d t \,   \dot{\tau}\left(\sin ^4\xi -1\right)  \, . 
\end{align}
The full D3 brane action becomes
\begin{align} \label{eq:SD3S5}
	S_{\text{D}3} & =  2 \pi ^2 T_3 L^3 \int d t  \left( -\sin ^3 \xi \sqrt{1+\frac{r^2}{L^2}-\frac{\dot{r}^2}{\frac{r^2}{L^2}+1}-\dot{\xi}^2- \dot{\tau}^2\cos ^2 \xi } +L \dot{\tau} \left( \sin ^4\xi    -1\right)\right).
\end{align}
The giant graviton solution is:
\begin{align} \label{eq:GGsolS5}
	\xi = \xi_0 \in \left[0,\, \frac{\pi}{2}\right], \quad \tau = \frac{t}{L}, \quad r =0\,.
\end{align}
The configuration with $r=0$ sits at the center of AdS$_5$. The parameter $\xi_0$ determines the size of the giant graviton. It is maximal for $\xi_0=\frac{\pi}{2}$; this particular configuration will be shown to play a central role in reproducing the dual supersymmetric index. There are several things to note at this stage. The coefficient of the $\dot{\tau}^2$ term in the D3-brane action is proportional to $\cos^2 \xi ={\rm e}^{- K}$, which vanishes as we approach the maximal giant configuration. Furthermore, the term proportional to $\dot{\tau}$ behaves as $\sim A_1 + A_2 \dot{\tau}$, for $A_{1,2}$ two finite constant values as we approach the maximal giant. This already helps build intuition towards the more generic geometries, where we can parametrize fluctuations by small deviations from ${\rm e}^{-K} =0$ and small deviations from $\dot{\tau}=\frac{1}{L}$. Let us now move on to the fluctuation analysis and try to understand how the geometry of the background and the wrapped cycle appears in the effective Lagrangian describing fluctuations.

\subsection{Fluctuation analysis for maximal giants on $S^5$}
We have been able to recover the action \eqref{eq:SD3S5} describing fluctuations about the maximal giant using the description of $S^5$ as a $U(1)$ fibration over $\mathbb{CP}^2$.  We will find it useful to study even the quantization of the single maximal giant graviton excitations using already known techniques in the context of the Quantum Hall effect. This will establish a nice connection with the approach used in \cite{Deddo:2024liu} to quantize the fluctuations of giant gravitons from bubbling geometries.
Let us consider small fluctuations around the giant graviton solution \eqref{eq:GGsolS5}:
\begin{align} \label{eq:fluctuationsGG}
	\xi(t) & = \xi_0 +  \delta \xi(t)\,, \quad \quad  \tau(t) = \frac{\omega}{L} t-  \delta \tau(t)\,,
\end{align}
where $\delta \xi(t)$ and $\delta \tau(t)$ are small fluctuations of the same order and $\omega$ is a parameter for the angular velocity that is equal to $1$ in the case of the giant graviton but we keep arbitrary for later purposes. The full set of fluctuations around the giant graviton in AdS$_5\times S^5$ is known \cite{Beccaria:2024vfx,Das:2000st,Gautason:2024nru,Deddo:2025lfm}, the purpose of our explicit computation is to track the role of the larger K\"ahler structure. Using the convenient set of variables $ \rho(t)  = \frac{\pi}{2} -\delta \xi(t), \quad \dot{\varphi}(t)  = \frac{\omega}{L} - \delta \dot{\tau}(t)\,,$ and the fact that the tension of D3 branes is  $T_3 L^3 = \frac{1}{2 \pi^2} \frac{N}{L}$, we can analyse the quadratic fluctuations around the maximal giant ($\xi_0 =\frac{\pi}{2}$). From \eqref{eq:Z3b}, we see that the constraint ensuring supersymmetry is slightly violated by the small fluctuations:
\begin{align} \label{eq:AbelianGauss}
	\cos\left(\frac{\pi}{2} - \delta \xi(t)\right) & \simeq \delta \xi(t) \neq 0 \,.
\end{align}
Even though we are only considering small fluctuations, it is clear that we cannot exactly satisfy the constraint $\delta \xi(t)=0$ while keeping non-trivial fluctuations on. We then introduce a new degree of freedom $\delta b(t)$ in such a way that both $b$ and $\cos\left(\frac{\pi}{2} - \delta \xi(t)\right)$ are deformed simultaneously:
\begin{align} \label{eq:deltab}
	b+  \delta b(t) & = \cos\left(\xi_0 - \delta \xi(t)\right) \Rightarrow \quad \delta b(t) = \delta \xi (t) \, \quad \text{for} \quad \xi_0 =\frac{\pi}{2}\,.
\end{align}
The idea now is to find a dynamical description of this new degree of freedom such that its equation of motion enforces the constraint $\delta b(t)=\delta \xi(t)$. Another way to interpret this is to think of \eqref{eq:AbelianGauss} as an Abelian version of the fact that the Gauss Law arising in the Quantum Hall effect described by a non-commutative matrix model \cite{Polychronakos:2001mi,Dai:2005hh}, cannot be satisfied exactly by finite size matrices. In this context it becomes necessary to introduce the so called edge modes to restore the conservation of charge in the system. Indeed, when considering the multiple giants excitations, we will study a matrix version of this statement and we will define an appropriate generalization of $\delta b(t)$ that we will call edge modes.

For the moment, let us try to choose appropriate variables to construct the Lagrangian describing the fluctuations. To do so, we can expand \eqref{eq:SD3S5}  up to second order in $\rho(t)$ and $\varphi(t)$ to obtain the following Lagrangian describing fluctuations
\begin{align}\label{eq:LD3polar}
	\mathcal{L}_{\text{D3}} & = \frac{N}{L}\left(\frac{1}{2} L^2\dot{\rho}^2+\frac{1}{2} L^2\rho^2 \dot{\varphi}^2+L\rho^2 \dot{\varphi}-1 \right) \,.
\end{align}
Interpreting $\rho, \varphi$ as polar coordinates on a plane, we can introduce better behave Cartesian coordinates following $\rho = \sqrt{\frac{x^2 + y^2}{L^2}}$ and $\varphi = \arctan(\frac{y}{x})$, then we can reduce \eqref{eq:LD3polar} to the following Lagrangian:
\begin{align} \label{eq:LanLagrangian0}
	\begin{split}   \mathcal{L}_{\text{D3}}& =\frac{N}{L} \left(\frac{1}{2} \left(\dot{x}^2+\dot{y}^2\right)+ \frac{2- \omega}{L} \left(x \dot{y}-y \dot{ x}\right)+\frac{(\omega-1)(\omega-3)}{2L^2}\left(x^2 +y^2\right)\right)-\frac{N}{L}\, .
	\end{split}
\end{align}
The above Lagrangian describes a charged particle moving on a plane under the action of a uniform transverse magnetic field and a quadratic potential. The constant term accounts for the contribution of the solenoidal term in \cite{Eleftheriou:2023jxr} and we interpret it here as a contribution from the vacuum energy. 

In practice the giant graviton solution sits at $\omega=1$, for which the quadratic potential vanishes. The reason to keep it arbitrary is twofold: (i) it manifestly shows the link between the strength of the magnetic field and the strength of the quadratic potential that will be important when studying the ground state and (ii) this term also appears naturally in the less supersymmetric backgrounds, specifically in the case of giant gravitons on AdS$_5\times T^{1,1}$. The Lagrangian we obtained corresponds to the Fock-Darwin system \cite{fock1928bemerkung,darwin1931diamagnetism}, which is extensively used by the condensed matter community, especially in the context of quantum dots, where magnetic confinement is combined with a parabolic potential\footnote{See also \cite{Drigho-Filho:2017bph} for more modern discussions on this system.}. 
Absorbing the factor of $L$ in the definition of the Lagrangian and rescaling the coordinates $(x, y, t) \rightarrow \left(\frac{L x}{\sqrt{N}}\,, \frac{L y}{\sqrt{N}}, \frac{t}{L}\right)$, we obtain:
\begin{align} \label{eq:Laglan0}
	\begin{split}
		\mathcal{L}_{\text{FD}}& =\frac{M}{2} \left(\dot{x}^2+\dot{y}^2\right)+ \frac{B}{2} \left(x \dot{y}-y \dot{ x}\right)+\frac{\Omega^2}{2}(x^2 +y^2)-N\,, \\
		B& =2(2 - \omega), \quad   \Omega^2= (\omega-1)(\omega-3)=-\frac{(\omega_c-2)(\omega_c+2)}{4}\, , \quad M=1\,.
	\end{split}
\end{align}
The canonical Hamiltonian associated with \eqref{eq:Laglan0} is the standard Fock-Darwin Hamiltonian,
\begin{align}
	\mathcal{H}_{\text{FD}}
	&=
	\mathcal{E}_0
	+\frac{1}{2M}\left(p_x^2+p_y^2\right)
	+\frac{M}{2}\left[\left(\frac{\omega_c}{2}\right)^2+\Omega^2\right]
	\left(x^2+y^2\right)
	-\frac{\omega_c}{2}\left(xp_y-yp_x\right),
	\label{eq:HFD}
\end{align}
where $\omega_c=B/M$ and $\mathcal{E}_0$ denotes the constant vacuum-energy shift. This is the Hamiltonian of a charged particle in a uniform magnetic field, with an additional harmonic confinement whose strength is fixed by the same parameter $\omega$ that controls the angular velocity of the giant.

Introducing the oscillator operators that diagonalize the Fock-Darwin problem, the Hamiltonian can be written as
\begin{align}
	\hat H_{\text{FD}}
	&=
	\omega_+\left(a^\dagger a+\frac12\right)
	+
	\omega_-\left(b^\dagger b+\frac12\right)
	+
	\mathcal{E}_0,
	\\
	\omega_\pm
	&=
	\sqrt{\left(\frac{\omega_c}{2}\right)^2+\Omega^2}
	\pm
	\frac{\omega_c}{2}.
\end{align}
Equivalently, after measuring energies in units of
$
\sqrt{\left(\frac{\omega_c}{2}\right)^2+\Omega^2},
$
we obtain
\begin{align}
	\hat H_{\text{FD}}
	&=
	(1+\nu)a^\dagger a
	+
	(1-\nu)b^\dagger b
	+
	1
	+
	\mathcal{E}_0,
	\qquad
	\nu
	=
	\frac{\omega_c}{
		2\sqrt{\left(\frac{\omega_c}{2}\right)^2+\Omega^2}
	}.
\end{align}
The angular momentum operator is
\begin{align}\label{eq:Langular}
	\hat L
	=
	b^\dagger b-a^\dagger a .
\end{align}

It is important to note that the Fock-Darwin system involves two oscillator sectors. For nonzero $\Omega$, the quadratic potential lifts the Landau-level degeneracy, while for $\Omega=0$ one has $\nu=1$ and the usual Landau problem is recovered. Nevertheless, the angular momentum operator \eqref{eq:Langular}, which is the operator relevant for counting states in the protected sector, is independent of $\Omega$. This observation is useful for the general SE$_5$ case: although the fluctuation dynamics around maximal giants will again take a Fock-Darwin form, the protected counting is governed by the angular-momentum grading, equivalently by the degeneracy structure that survives in the lowest Landau-level limit.

\subsection{Supersymmetric Fock-Darwin system}
In this subsection, we will mainly follow \cite{Ivanov:2019rbe}.
To write the supersymmetric version of the Fock-Darwin system, it will be convenient to work in complex coordinates
\begin{align}\label{eq:zzbar}
	z & = \frac{x+ i y}{\sqrt{2 }}, \quad \bar{z} = \frac{x - i y}{\sqrt{2 }}\,, 
\end{align}

\begin{align}
	\mathcal{L}_{\text{FD}} & = M\dot{z} \dot{\bar{z}} + \frac{i M  \omega_c}{2}\left(z \dot{\bar{z}} - \bar{z} \dot{z}\right)-  M\Omega^2 z \bar{z}  \,.
\end{align} 
Defining $\kappa(z ,\, \bar{z})$ as the K\"ahler potential for the complex manifold parametrizing the fluctuations about the maximal giant. For the case of $S^5$,  $\kappa(z, \, \bar{z})=z \bar{z}$. The mass term can be absorbed in the definition of the Lagrangian in such a way that we can  rewrite:
\begin{align}
	\mathcal{L}_{\text{FD}} & = g_{z \bar{z}} \dot{z} \dot{\bar{z}} + i \frac{\omega_c}{2}\left(\partial_z \kappa \dot{z} - \partial_{\bar{z}}\kappa\dot{\bar{z}}\right) - \Omega^2g^{z \bar{z}}\partial_z \kappa \partial_{\bar{z}} \kappa\, .
\end{align}
The Hamiltonian is then:
\begin{align}
	\mathcal{H}_{\text{FD}}  & = g^{z \bar{z}}\left(p_{\bar{z}} p_z + \Omega^2 \partial_z \kappa \partial_{\bar{z}} \kappa\right)\,.
\end{align}
The supersymetric version of this system is obtained following \cite{Ivanov:2019rbe}:
\begin{align} \label{eq:LSFD}
	\mathcal{L}_{\text{SFD}}& =   g_{z \bar{z}} \dot{z} \dot{\bar{z}} + i \frac{\omega_c}{2}\left(\partial_z \kappa \dot{z} - \partial_{\bar{z}}\kappa\dot{\bar{z}}\right) - \Omega^2g^{z \bar{z}}\partial_z \kappa \partial_{\bar{z}} \kappa +\frac{i}{2}g_{z \bar{z}} \left(\eta^{\alpha}\dot{\bar{\eta}}_{\alpha}+\bar{\eta}_{\alpha} \dot{\eta}^{\alpha}\right) - \frac{\omega_c}{2}\,g_{z\bar{z}}\,\eta^{\alpha}\,\bar{\eta}{}_{\!\alpha} \, .
\end{align}
with Hamiltonian:
\begin{equation} \label{eq:SFD}
	\mathcal{H}_{\text{SFD}} =g^{z\bar{z}}\Big(p_z \,\bar{p}_{\bar{z}} + \Omega^{2}\,\partial_{z}\kappa\,\partial_{\bar{z}}\kappa\Big)
	+ \frac{\omega_c}{2}\,g_{z\bar{z}}\,\eta^{\alpha}\,\bar{\eta}{}_{\!\alpha}\,,
\end{equation}
where $\eta$ is a fermionic degree of freedom and the supersymmetric completion reduces to the standard Pauli term coupling the spin of the electron to the external magnetic field used in \cite{Eleftheriou:2023jxr, Eleftheriou:2025lac}.
The Lagrangian \eqref{eq:LSFD} is invariant under the following supersymmetric transformations:
\begin{align}
	\delta z & =  a \eta \,, \quad \quad \quad \quad \quad \quad \quad \,\, \delta \bar{z} = \bar{a} \bar{\eta}\, , \\
	\delta \eta & = - g_{z\bar{z}} \left(\dot{z}+ i \alpha \partial_{\bar{z}}\kappa\right), \quad \delta \bar{\eta} = - g_{z\bar{z}} \left(\dot{\bar{z}}- i \bar{\alpha} \partial_{z}\kappa\right)\, ,
\end{align}
with $\alpha \equiv \frac{2\Omega^2}{\omega_c}$:
\begin{align}
	\begin{split}
		\delta \mathcal{L}_{\text{SFD}} & = a g_{z \bar{z}} \dot{\bar{z}} \dot{\eta} + \bar{a} g_{z \bar{z}} \dot{z} \dot{\bar{\eta}} +i \frac{\omega_c}{2}\left(\bar{a} \bar{\eta} \dot{z}- a \eta \dot{\bar{z}}+ a \partial_z \kappa \dot{\eta}- \bar{a} \partial_{\bar{z}}\kappa \dot{\bar{\eta}}  \right) \\
		& - \Omega^2\left(\bar{a} \bar{\eta} \partial_{\bar{z}}\kappa + a \partial_{z}\kappa \eta\right)- \frac{\omega_c}{2}\left(\delta \eta \bar{\eta}- \eta \delta \bar{\eta}\right) \\
		& = i \frac{\omega_c}{2}\left(\bar{a} \bar{\eta} \dot{z}- a \eta \dot{\bar{z}}\right)+ \bar{\eta}\left(\frac{\omega_c}{2}\delta \eta - \bar{a} \Omega^2 \partial_{\bar{z}} \kappa\right) + \eta \left(\frac{\omega_c}{2} \delta \bar{\eta} - a \Omega^2 \partial_z\kappa\right) \,. 
	\end{split}
\end{align}
The logic of the localization argument can now be summarized as follows:

\begin{itemize}
	\item The supersymmetric completion \eqref{eq:LSFD} can be written, up to total derivatives, as a $\delta$-exact deformation,
	\begin{align}
		\mathcal{L}_{\text{SFD}}
		=
		\delta \mathcal{V}\,,
	\end{align}
	for an appropriate choice of gauge fermion $\mathcal{V}$. This makes the partition function insensitive to continuous deformations of the localization parameter.
	
	\item The bosonic part of the localization equations is obtained by setting the fermionic variations to zero. For the maximal giant this gives
	\begin{align}
		\dot z=0,
		\qquad
		\partial_{\bar z}\kappa=0\,,
	\end{align}
	whose regular solution is
	\begin{align}
		z=0\,.
	\end{align}
	Thus the fixed locus coincides with the maximal giant configuration.
	
	\item Quantum fluctuations around the localization locus are therefore captured by the Hilbert space of the corresponding lowest Landau-level problem. For $S^5$ this Hilbert space is the space of holomorphic wavefunctions on the transverse plane, while for a general SE$_5$ background it is replaced by the appropriate equivariant Hilbert space determined by the local K\"ahler  geometry of the cone.
	
	\item The protected contribution to the giant graviton expansion is obtained by taking the equivariant trace over this Hilbert space,
	\begin{align}
		\mathrm{Tr}_{\mathcal{H}_{\text{LLL}}}
		(-1)^F q^{\hat L},
	\end{align}
	or its refined version when additional flavor fugacities are present. Localization ensures that this index is independent of the details of the deformation and depends only on the protected grading data.
\end{itemize}

Rather than reproducing the localization computation leading to the giant-graviton contribution to the index, we emphasize the two structural ingredients that will generalize to arbitrary SE$_5$ backgrounds. First, the angular momentum operator \eqref{eq:Langular}, which grades the protected Hilbert space, is insensitive to the strength of the confining potential and survives the lowest-Landau-level limit $|\omega_c|\rightarrow\infty$. Second, the ambient geometry enters the fluctuation problem only through the K\"ahler potential $\kappa(z,\bar z)$, suggesting that in the general SE$_5$ case the transverse K\"ahler geometry completely determines the protected sector relevant for the giant graviton expansion.

\subsection{The giant graviton index from the Quantum Hall effect}
We will now exploit the fact that the $\frac{1}{2}$-BPS sector can be captured by the lowest Landau level of the system \eqref{eq:Laglan0} \cite{Eleftheriou:2023jxr, Eleftheriou:2025lac}. 
Ignoring for now that $\omega_c$ is a fixed parameter, we can project the system into the lowest lying Landau level by taking the limit $|\omega_c| \rightarrow \infty$. The constant term is not affected by the LLL projection; we keep it as the classical energy of the wrapped D3-brane, which later produces the expected factor $q^N$ in the giant graviton expansion. We then have:
\begin{align}
	\mathcal{L}_{\text{LLL}} & = i\frac{M \omega_c}{2}\left(z \dot{\bar{z}}- \bar{z} \dot{z}\right)- M\Omega^2 z \bar{z}  - N\,.
\end{align}
Up to a total derivative, the first-order term may be written as $-iM\omega_c \bar z \dot z$.
Defining canonical momenta:
\begin{align}
	p_z & = \frac{\partial \mathcal{L}_{\text{Lan}}
	}{\partial \dot{z}} =  - \frac{i M \omega_c}{2} \bar{z}\, , \quad p_{\bar{z}} =  \frac{\partial \mathcal{L}_{\text{Lan}}
	}{\partial \dot{\bar{z}}} = \frac{i M \omega_c}{2} z\, .
\end{align}
The Hamiltonian of this problem is
\begin{align}
	\mathcal{H}_{\text{LLL}} & = N+ 
	M\Omega^2 z \bar{z} \,.
\end{align}
In the discussion so far we have ignored the new type of degree of freedom $\delta b(t)$ defined in \eqref{eq:deltab}. This is a good point to return to it. We would like to define a theory describing fluctuations characterized by $\delta \xi^2 \sim z \bar{z}$, and $\delta b$, in a way that one of the equations of motion of the theory ends up identifying these two quantities. One interesting way to do this, that will turn out instrumental in the multi-giant section, is to introduce a background gauge field $a_0$, such that, when we integrate it out, we obtain $\delta \xi^2 \sim 1- \bar{\chi} \chi$. We then turn on a non-dynamical gauge field $a_0$ that couples to both the $z$ and $\chi$ fields and construct the Lagrangian as follows
\begin{align} \label{eq:LU1}
	\mathcal{L}_{U(1)} & = - i M \omega_c \bar{z} D_t z - M \Omega^2 z \bar{z} -N - i M \omega_c \bar{\chi}D_t \chi+   M \omega_c a_0 
\end{align}
\begin{align}
	D_t z & = \dot{z} - i  a_0 z, \, \quad D_{t} \chi =\dot{\chi}- i  a_0 \chi \,.
\end{align}
We see that the equations of motion of $a_0$ generate the constraint:
\begin{align}
	z \bar{z} + \bar{\chi} \chi & = 1\,.
\end{align}
The non-Abelian generalization required to describe $m$ giants becomes  immediate and makes contact with the matrix model descriptions well studied in the literature of Quantum Hall effect \cite{Polychronakos:2001mi, Dorey:2016mxm, Dorey:2016hoj}.

\subsection{Coincident giants and the superconformal index}
Let us now promote the coordinates describing the fluctuations to be Hermitian $m \times m$ matrices, transforming in the adjoint representation of the group $U(m)$.
Setting the mass $M=1$ we can directly generalize the Lagrangian \eqref{eq:LU1} as follows:
\begin{align} \label{eq:QHMM0}
\mathcal{L}_{U(m)}
=
\operatorname{Tr}\left(
-i\omega_c Z^\dagger D_t Z
-\Omega^2 Z Z^\dagger
-N\mathbbm{1}_m
+\omega_c A_0
\right)
-i\omega_c \Phi^\dagger D_t\Phi .\,,
\end{align}
where $z \rightarrow Z_{ab} \in U(m), \, \chi \rightarrow \Phi_a, a_0 \rightarrow A_0$ such that
\begin{align}
	\begin{split}
		Z& \rightarrow Z \, U\, U^{\dagger} , \quad \Phi \rightarrow U \, \Phi, \, \\
		D_t Z & =\partial_t Z - i [A_0, \, Z], \quad D_t \Phi = \partial_t 
		\Phi - i A_0 \Phi \,.
	\end{split}
\end{align}
Then the matrix version of the giant graviton constraint would be implemented as the Gauss Law for the gauge background field $A_0$:
\begin{align}
	[Z, \, Z^{\dagger}]  +  \Phi \Phi^{\dagger} & = \mathbbm{1}_m\, .
\end{align}
The partition function of a system similar to this one has been analyzed in detail in \cite{Dorey:2016hoj}, with the difference that in their case they included a chemical potential associated with the edge modes $\Phi$.

The Hamiltonian for this system is:
\begin{align}
	\mathcal{H}_{U(m)} & = m N+
	\Omega^2 \text{Tr}(Z^{\dagger} Z) \,. 
\end{align}

The partition function of this system is obtained by tracing ${\rm e}^{- \gamma \mathcal{H}_{U(m)}}$ over the Hilbert space of $Z^{\dagger}$ excitations, with $\gamma$ being the length of the thermal circle. The partition function can be further refined in a grand-canonical ensemble turning on a chemical potential for the $\Phi^{\dagger}$ excitations.
\begin{align}
	\mathcal{Z}(q, x)& ={\rm e}^{- \gamma m N}
	\text{Tr}_{\text{QH}} \,{\rm e}^{- \gamma ( \frac{\Omega^2}{|\omega_c|} n- \mu j)} , \quad q \equiv {\rm e}^{- \gamma \frac{\Omega^2}{|\omega_c|}}, \quad x \equiv {\rm e}^{\gamma \mu}\,.
\end{align}
Some comments on the definition of the fugacity are in order: The partition function is computed as a trace over the Hilbert space of lowest Landau level excitations in the presence of a supersymmetric background. The degeneracy of the Landau problem is lifted by introducing a harmonic regulator, which should be understood as a Q-exact deformation of the Hamiltonian. Such deformations modify the spectrum away from the BPS locus but leave invariant the protected index.

In this sense, the parameter $	\frac{\Omega^2}{|\omega_c|}$
 does not represent a physical coupling, but a choice of supersymmetric grading that determines how the BPS cohomology is embedded into the full Hilbert space. 
\begin{align} \label{eq:qReg}
	\frac{\Omega^2}{(-\omega_c)}= 1\, ,
\end{align}
which yields a natural regularization $q ={\rm e}^{-\gamma}$.
An appropriate complex shift of the chemical potential $\mu$, allows us to insert the fermion number operator, thus effectively turning the partition function into the index that we want to compute. Specifically, we shift the chemical potential $\mu$ by $\frac{i \pi}{\gamma}$, hence $x^j \rightarrow x^{\gamma \mu (j+ \frac{i \pi}{\gamma \mu}j)}=(-1)^j q^{-j}$. Moreover, to connect with the $\frac{1}{2}$-BPS sector of giant graviton excitations, we identify the chemical potentials of the Hamiltonian and the angular momentum $J= -R$. We then obtain:
\begin{align}
	\mathcal{I}_{U(m)} &  = \mathcal{Z}(q, -q) = {\rm e}^{- \gamma m N}
	\text{Tr}_{\frac{1}{2}\text{-BPS}} \, (-1)^Fq^{n -j} ={\rm e}^{- \gamma m N}
	\text{Tr}_{\frac{1}{2}\text{-BPS}} \, (-1)^Fq^{ r}\,.
\end{align}
Using the result reviewed in the appendix, we evaluate equation \eqref{eq:ZQHE} in $x =-q$ we obtain
\begin{align}
	\mathcal{I}_{U(m)} & = (-1)^m q^{m N}q^{\frac{m(m+1)}{2}} \prod_{j=1}^m \frac{1}{1- q^j}\,.
\end{align}
The above expression $\mathcal{I}_{U(m)}$ is precisely the type of giant graviton index, introduced in \eqref{Eq:GGExpansions}, leading to a precise giant graviton expansion of the index. 

The goal now is to analyze the remaining SE$_5$  manifolds using the same framework. In particular, we would like to identify the massless excitations sector that will yield the analogoue of the Landau problem whose quantization accounts for the index of giant gravitons.

		\section{The giant graviton expansion in AdS$_5\times {\rm SE}_5$: a general discussion} \label{sec:GeneralKE}
	In this section we first review the field theory side of the giant graviton expansion of the superconformal index defined on the worldvolume theories of a stack of $N$ D3-branes probing the tip of toric Calabi-Yau cones whose near-horizon geometry is AdS$_5\times$SE$_5$ (in practice we set SE$_5 =Y^{p,q}$). We also discuss the generalization to this framework of the fluctuation analysis discussed in the previous section. The goal is to identify the extent to which the Hilbert space of quantum fluctuations associated with BPS D3 brane configurations entering the giant graviton expansion can be captured by appropriate generalizations of the Landau problem (and the Quantum Hall matrix model) appearing in the $S^5$ case. 
	\subsection{The field theory side}
	
	The superconformal index \cite{Romelsberger:2005eg, Romelsberger:2007ec, Kinney:2005ej} of $\mathcal{N} =1$ toric quiver theories on $S^1 \times S^3$ can be written as: 
	\begin{equation}
		\mathcal{I}_N(\vec{y}; \vec{q}) = \text{Tr}_{\mathcal{H}(S^1\times S^3)}\left[\left(-1\right)^{F}y_1^{J_1 + \fft{R}{2}} y_2^{J_2 + \fft{R}{2}}\prod_{I=1}^d q_I^{Q_I}\right], \label{Eq:TheSCI}
	\end{equation}
	where $Q_I$ are the flavor charges with fugacities $q_I$. The combination $J_{1,2}+\frac{R}{2}$, where $J_{1,2}$ are the angular momenta on $S^3$ and $R$ is the R-charge, commute with the supercharge that annihilates the BPS states counted by the index. The fugacities satisfy the constraint $\prod_{I=1}^d q_I = y_1 y_2$. Then \eqref{Eq:TheSCI} has been written as:
	\begin{align}\label{Eq:SCI-general}
		\mathcal{I}_{N}(\vec{q})& = \mathcal{I}_{\infty} \left(\sum_{m_1=0}^{\infty} \cdots \sum_{m_d=0}^{\infty} q_1^{m_1 N} \cdots q_d^{m_d N} \mathcal{J}_{m_1, \cdots, m_d}(\vec{q})\right)\,,
	\end{align}
	where, in the case of toric SE$_5$, $d$ is the number of external vertices of the toric diagram and $m_1, \cdots, m_d$ correspond to the wrapping numbers of the giant gravitons around supersymmetric cycles inside SE$_5$. Furthermore we have defined the fugacities $\vec{q}=\{q_1, \cdots, q_d\}$. Each set of wrapping numbers $\{m_1, \cdots, m_d\}$ characterizes a $U(m_1)\times\cdots \times U(m_d)$ quiver gauge theory realized by this particular configuration of D3-branes. The function $\mathcal{J}_{m_1, \cdots, m_d}(\vec{q})$ is such that $\mathcal{J}_{0, \cdots, 0}(\vec{q})=1$ and it has been described, at least formally, as the index in this field theory.
	Another convenient presentation of \eqref{Eq:SCI-general} can be obtained by redefining the fugacities as follows:
	\begin{align}
		q_I & = \widetilde{q} \prod_{b=1}^{d-3}\zeta_{b}^{B_{b,I}} \prod_{\alpha=1}^2u_{\alpha}^{F_{\alpha, I}}\,.
	\end{align}
We are interested in unrefined limits of the index that isolate the protected giant-graviton sector; for the moment, we set $u_{1}=u_2 =1$.
	\begin{align} \label{eq:Iqtilde}
		\mathcal{I}_{N}(\vec{q})& = \mathcal{I}_{\infty} \left(\sum_{m_1=0}^{\infty} \cdots \sum_{m_d=0}^{\infty} \widetilde{q}^{m N}  \prod_{b=1}^{d-3}\zeta_b^{N B_b} \mathcal{J}_{m_1, \cdots, m_d}(\vec{q})\right)
	\end{align}
	where $m = \sum_{I=1}^d m_I$ is the total wrapping number, $B_b=\sum_{I=1}^d B_{b,I} m_I$ is the total baryonic charge with associated fugacity $\zeta_b$. For the $Y^{p,q}$, we have simply $d=4$, therefore there is only one baryonic charge, associated with the homology class of the wrapped three-cycle.
	For $y_1=y_2 \equiv q$, we have $\prod_{I=1}^4 q_I = \widetilde{q}^4 = q^2$, where $q$ is the fugacity associated with the $R$-charge. This is going to be important later to determine the appropriate fugacity to perform the series expansion.
	
	A simplified setup is obtained by considering contributions $\mathcal{J}_{I,\,m}(q_I)= \mathcal{J}_{0, \cdots, m, \cdots, 0}(\vec{q})$ associated with giant gravitons wrapping one of the 3-cycles labeled by $I=1,\cdots, d$. After an appropriate change of variables  \cite{Imamura:2022aua,Gaiotto:2021xce} and being careful with the pole structure of the functions involved, it was found that the index has the form: 
	\begin{align} \label{eq:JIthCycle}
		\mathcal{I}_{N}(q) & = \mathcal{I}_{\infty} \left(\sum_{m=0}^{\infty} q_I^{m N} \mathcal{J}_{I, m}(q_I)\right)\,.
	\end{align} 
	Notice that in this case, no refinement via the baryonic charge is necessary as one is wrapping a single type of cycle multiple times, so we can safely set $\zeta=1$. In configurations wrapping combinations of several types of 3-cycles, that is, at least two $m_{I}, m_J \neq 0$, it would become important to refine by baryonic charges. In fact, the quantization process would have to be done sector by sector. The only such configuration we will analyze is the zero baryonic charge sector, which does not require us to turn on $\zeta$.  We shall comment further on these configurations below.


	\subsection{The gravity side}
	On the gravity side, we expect that the quantization of fluctuations about the  $m$  giants wrapping the $I$-th supersymmetric cycle can reproduce the $m$-th term in \eqref{eq:JIthCycle}. Let us make a clarification about terminology: strictly speaking, the D3-branes wrapping one of the above-mentioned supersymmetric 3-cycles are not the giant gravitons as originally described in \cite{McGreevy:2000cw}, since the giant gravitons wrapped trivial 3-cycles, such that they were stabilized dynamically. In contrast, the D3-branes wrapping the topologically non-trivial cycles $\Sigma_I$ cannot shrink to zero size and are therefore topologically stable. The above expansions are nevertheless called giant graviton expansions because of their origin in the case of the superconformal index $\mathcal{N}=4$ SYM, where the 3-cycles wrapped inside $S^5$ are always topologically trivial. There is, however, a class of D3-brane configurations that fall within the same category as the $S^5$ giant gravitons, corresponding to the situation in which the branes wrap combinations of 3-cycles that, when combined, become trivial \cite{Balasubramanian:2001nh, Beasley:2002xv}. We will also study the quantization of these configurations corresponding to the zero baryonic charge sector.
	
	For concreteness, let us focus on the toric case, and more specifically on the class of SE$_5$ known as $Y^{p,q}$, which have been extensively studied in the literature \cite{Gauntlett:2004yd, Martelli:2004wu} and their metrics are known. Topologically, we have that $Y^{p,q} \cong S^2 \times S^3 $. The four-dimensional basis $B_4$  is an axially squashed trivial $S^2$ bundle over $S^2$ \cite{Martelli:2004wu},=. Therefore $B_4\cong S^2 \times S^2$.
	Locally, the metric on $Y^{p,q}$ can be given by:
	\begin{align} 
		ds_{\text{SE}_5}^2 & = \frac{1 - c y}{6}\left(d \theta^2 + \sin^2\theta d \phi^2\right) + \frac{1}{W(y) Q(y)}d y^2 + \frac{Q(y)}{9}\left[d \psi - \cos \theta d \phi\right]^2 \label{eq:ThemetricYpq} \\
		\nonumber
		&+ W(y) \left[d \alpha +\frac{a c - 2 y + c y^2}{6( a - y^2)}\left(d \psi - \cos \theta d \phi\right)\right]^2 \\
		ds_{\text{SE}_5}^2  & = ds^2\left(B_4\right) + W(y)\left(d \alpha  +A\right)^2\,,  \label{eq:B4}\\ 
		A & = \frac{a c - 2 y + c y^2}{6 (a - y^2)} \left(d \psi - \cos \theta d \phi\right),
	\end{align}
	where:
	\begin{align}
		W(y) & = \frac{2 \left(a  - y^2\right)}{1 - c y} \,,\\\label{eq:Qy}
		Q(y) & = \frac{a - 3 y^2 + 2 c y^3}{a - y^2}. 
	\end{align}
	The variable $y$ has to obey the following constraints: $y<1, \hspace{1.5mm} a - y^2 >0, \hspace{1.5mm} W(y)>0, \hspace{1.5mm} Q(y)\geq 0$. We define $y_1, y_2, y_3$ such that
	\begin{align}
		(y -y_1)(y- y_2)(y - y_3) & = a- 3 y^2+ 2 y^3, \label{eq:yroots}
	\end{align}
	where for $0< a <1$, we have $y_1 <0$ and $y_{2,3} >0$. Then, for $y_1< y < y_2$, we have $0\leq \theta \leq \pi, \hspace{1.5mm} 0 \leq \phi \leq 2 \pi, \hspace{1.5mm} 0 \leq \psi \leq 2 \pi$.  
	Note that for $c\neq 0$ we can always set it equal to one, as we did in \eqref{eq:yroots}, whereas if $c=0$, then \eqref{eq:ThemetricYpq} is the standard homogeneous metric for $T^{1,1}$.
	The BPS giant gravitons that we are interested in are described in \cite{Mikhailov:2000ya} as the intersection locus between $Y^{p,q}$ and holomorphic hypersurfaces in the Calabi-Yau cone $C(Y^{p,q})$ whose base is $Y^{p,q}$. To make this statement more precise, we need to define appropriate coordinates on the cone. It is possible to define $C(Y^{p,q})$ as the K\"ahler quotient $\mathbb{C}^4//U(1)$ defined by the moment map constraint:
	\begin{align}\label{eq:GLSM}
		\sum_{I=1}^4 B^I |z_I|^2 & = 0\,, \quad \{B^I\}=\{p,\, p, -p+q, -p-q\}\,,
	\end{align}
modulo the $U(1)$ action $z_I\rightarrow \lambda^{B^I} z_I, \, \lambda \in \mathbb{C}^*$ on the complex coordinates $z_I$ of $\mathbb{C}^4$.
	For the purpose of studying the maximal configurations, it will be useful to choose a set of variables $w_{J=1,\cdots,4}$ that are left invariant by the $U(1)$ action: $w_J = \prod_{I}z_I^{a_{JI}}$ for appropriate values of $\{a_{JI}\}$ (see for example \cite{Forcella:2008bb}). 
	The holomorphic hypersurface defining BPS giant gravitons at an initial time can be defined as:
	\begin{align} \label{eq:FMik}
		F(w_J) & = 0, \, \text{for} \quad J=1,\cdots, 4\,.
	\end{align}
	The variables $w_J$, when interpreted holographically, will correspond to combinations of the fields with vanishing baryonic charge. From the point of view of the SE$_5$ basis, the maximal giant graviton wraps BPS 3-dimensional sub-manifolds inside $Y^{p,q}$, specifically they wrap individually non-trivial cycles that, after being joined together, are trivial \cite{Balasubramanian:2001nh}. Since the total baryonic charge is zero, the wrapped combination of cycles is contractible and such D3-brane configurations are kept dynamically stabilized, namely through the angular momentum associated with rotation and the coupling to $C_4$. In this sense they are on the same footing as the giant gravitons studied on $S^5$.
	
	Supersymmetric 3-cycles $\Sigma_I \cong S^3/\mathbb{Z}^{k_I}, \, k_I=\{p,p,p-q, p-q\}$ are the bases of cones defined by $z_I=0$ (complex divisors of $C(Y^{p,q})$). 
	From the metric \eqref{eq:ThemetricYpq}, this statement can be rephrased roughly by saying that the maximal giants exist at $\theta=0,\pi$ (North and South poles of $S^2$) and at $y=y_1,y_2$ where $y_i$ are the roots of $Q(y)$ defined in \eqref{eq:Qy}: effectively $y_1$ and $y_2$ and the north and south poles of the 2-cycle underlying $\Sigma_I$. 
	
	Consider the monomial 
	\begin{align}\label{eq:Wib}
		w_J &= \prod_{I}z_I^{a_{JI}} = b, \quad \text{for some}\,\quad J=1, \cdots, 4\, ,
	\end{align}
For exponents $a_{JI}$ such that $w_J$ has vanishing baryonic charge, the equation \eqref{eq:Wib} defines a Mikhailov-type giant graviton configuration, since it is a particular case of \eqref{eq:FMik}. In the maximal limit $b=0$, the hypersurface degenerates into a reducible divisor,
\begin{align}
	w_J=0
	\qquad \Longleftrightarrow \qquad
	\bigcup_{I\,:\,a_{JI}>0}\{z_I=0\}\,,
\end{align}
with multiplicities determined by the exponents $a_{JI}$. Thus the maximal configuration is a union of supersymmetric cycles $\Sigma_I$ whose total baryonic charge vanishes.

For nonzero but small $b$, the components of this reducible divisor are smoothed into a single holomorphic hypersurface. Equivalently, a small fluctuation $\delta b(t)$ recombines the elementary components while preserving the holomorphic constraint \eqref{eq:Wib}. This is the analogue, in the present geometric setting, of the additional boundary or edge degree of freedom introduced in the quantum-Hall description of the $S^5$ giant.

Suppose, for example, that we focus on a maximal component localized at one of the degeneration loci $y=y_i$. Small fluctuations can be parametrized locally by
\begin{align}
	y=y_i+\delta y_i(t)\,,
\end{align}
or, equivalently, through the deformation parameter
\begin{align}
	w_J(y_i+\delta y_i(t))=\delta b(t)\,.
\end{align}
Specifying the supersymmetric embeddings of submaximal configurations directly in the coordinates of \eqref{eq:ThemetricYpq} is cumbersome. For explicit manipulations, it is more useful to exploit the transverse K\"ahler-Einstein structure of the SE$_5$. As in the $S^5$ example, the departure from the maximal giant can then be encoded locally by the behavior of the transverse K\"ahler potential, or equivalently by a parameter of the form ${\rm e}^{-K}$.

	\subsection{Maximal giants from the transverse K\"ahler-Einstein structure}
	The generic SE$_5$ manifold is a $U(1)$ fibration over a K\"ahler-Einstein base. 
	Let $K(\zeta,\bar{\zeta})$ be the K\"ahler potential of the 4d K\"ahler-Einstein base. Then we can write the AdS$_5\times$SE$_5$ type IIB background as: 
	
	\begin{align} \label{eq:SEmetricK}
		\begin{split}
			ds^2&= ds^2_{{\rm AdS}_5}+ds^2_{{\rm SE}_5},  \\
			ds^2_{{\rm SE}_5}&= \left(d\beta +\frac{\i}{2}(K_{,i}d\zeta^i - K_{,\bar{i}}d\bar{\zeta}^i)\right)^2+ K_{, i \bar{j}}d\zeta^i d\bar{\zeta}^j \, ,  \\
			F_5&= {\cal F}+ \ast {\cal F},  \\
			{\cal F}&=  {\rm vol}(\text{SE}_5)=\frac{1}{2}d\beta \wedge J \wedge J, \, \quad J= d \mathcal{A} = i K_{,\, i \bar{j}} d \zeta^i\wedge d \bar{\zeta}^{j}\,.
		\end{split}
	\end{align}
	The flux is chosen proportional to the volume form  of the internal manifold, which is written in terms of the $2$-form $J = d \mathcal{A}$. Here $\mathcal{A}$ is the $1$-form connection:
	\begin{align}
		\mathcal{A} & = \omega_i d \zeta^i + \omega_{\bar{i}} d \bar{\zeta}^i = \frac{i}{2}\left(K_{,i}d\zeta^i - K_{,\bar{i}}d\bar{\zeta}^i\right). \,
	\end{align}
	We consider the same parametrization of AdS$_5$ as in \eqref{Eq:AdS5}.
	The full metric can be split as follows:

	\begin{eqnarray}
		G_{MN}d X^M d X^N &=& g_{\mu \nu} d x^{\mu} d  x^{\nu} + h_{\beta \beta} d\beta^2 + 2 h_{\beta i} d\beta d\zeta^i +  2 h_{\beta \bar i} d\beta d\bar {\zeta}^{\bar i} + h_{ij} d\zeta^i d\zeta^j  \nonumber  \\
		&+&  h_{i \bar j } d\zeta^i d \bar  \zeta^{\bar j}+ h_{\bar i \bar j} d\bar \zeta^{\bar i} d\bar \zeta^{\bar j} \,,
	\end{eqnarray} with
	\begin{eqnarray}
		h_{\beta \beta} &=& 1\,,  \,\, h_{\beta i} =   \omega_i\,, \,\, h_{\beta \bar i} =     \omega_{\bar i}\,,  \,\, h_{ij} =h_{ji} =     \omega_i \omega_j \,,    \nonumber \\ 
		h_{\bar i \bar j} &=& h_{\bar j \bar i} =      \omega_{\bar i}  \omega_{\bar j} \,, \,\, h_{i \bar j} =  2\omega_i  \omega_{\bar j} +  K_{,i \bar j}\,.
	\end{eqnarray}
	Here the indices $i, \bar{i} =1,2$  and the four remaining coordinates are numbered $\alpha, \beta=6\,,\ldots, 9$. 
	The induced metric on the D-brane worldvolume is:
	\begin{align}
		\gamma_{a b} & = G_{MN} \frac{\partial X^M}{\partial \sigma^{a}} \frac{\partial X^N}{\partial \sigma^{b}} \, \quad  \text{where} \,\, a, b =0\,,\ldots,\,3 \,, \, \quad M, N = 0\,,\ldots,\,9\,.
	\end{align}
	Consider now an embedding of the form 
	\begin{eqnarray} \label{eq:embeddingeqs}
		t = \sigma^0, \,\beta =\beta(\sigma^0), \, r = r(\sigma^0)\,\, \textrm{and}\,\, \zeta^{i} = \zeta^i(\sigma^0,\sigma^n), \,\, \textrm{where} \,\, n=1,2,3\,.
	\end{eqnarray}
We restrict to embeddings for which the transverse K\"ahler potential is constant along the spatial worldvolume directions and depends only on time. Therefore $K(\zeta^i, \, \bar{\zeta}^i) = K(\sigma^0)$. Then, denoting  the derivatives with respect to $\sigma^0$ by a dot, we have
	\begin{eqnarray}
		\gamma_{00}  &= & g_{00} + g_{11} \dot{r}^2 + h_{\beta \beta}\dot{\beta}^2 + 2 \dot{\beta}\left(h_{\beta i} \dot{\zeta}^i + h_{\beta \bar{i}} \dot{\bar{\zeta}}^i\right) + 2 h_{i \bar{j}} \dot{\zeta}^i \dot{\bar{\zeta}}^j + h_{ij} \dot{\zeta}^i \dot{\zeta}^j  +h_{\bar{i} \bar{j}} \dot{\bar{\zeta}}^i \dot{\bar{\zeta}}^j \,, \\
		\gamma_{0n} &=& \dot{\beta} \left( h_{\beta i  } \frac{\partial \zeta^i}{\partial \sigma^n}  +  h_{\beta \bar{i}  } \frac{\partial \bar{\zeta}^i}{\partial \sigma^n}  \right) + 2 h_{i \bar{j}} \frac{\partial{{\zeta}^i}}{\partial \sigma^{(0}} \frac{\partial \bar{\zeta}^{\bar j}}{\partial \sigma^{n)}} + h_{i j}  \frac{\partial{{\zeta}^i}}{\partial \sigma^{(0}} \frac{\partial \zeta^j}{\partial \sigma^{n)}}  + h_{\bar{i} \bar{j}} \frac{\partial{{\bar \zeta}^{\bar i}}}{\partial \sigma^{(0}} \frac{\partial \bar{\zeta}^j}{\partial \sigma^{n)}} \,, \\
		\gamma_{mn} &=& 2 h_{i \bar{j}} \frac{\partial \zeta^i}{\partial \sigma^{(m}} \frac{\partial \bar{\zeta}^{\bar j}}{\partial \sigma^{n)}} +   h_{i j}  \frac{\partial \zeta^i}{\partial \sigma^{(m}}\frac{\partial \zeta^j}{\partial \sigma^{n)}}  + h_{\bar{i} \bar{j}} \frac{\partial \bar{\zeta}^i}{\partial \sigma^{(m}}\frac{\partial \bar{\zeta}^{\bar j}}{\partial \sigma^{n)}}   \,,
	\end{eqnarray}  where $a_{(1} b_{2)} = \frac{1}{2 } \left(a_1 b_2 + a_2 b_1\right)$, i.e., symmetrization here and henceforth is always with weight 1.
	
	When evaluating $\det \gamma$, we organize the determinant as a quadratic polynomial in $\dot\beta$. This isolates the terms that will become the effective kinetic and magnetic couplings for fluctuations around the maximal giant. Schematically, we have:
	\begin{align}
		\det \gamma & = \left(f_1 \dot{\beta}^2 + f_2 \dot{\beta} + f_3\right)\det \gamma_{mn}  \,,
	\end{align}
	where 
	\begin{align}
		\begin{split}
			f_1 & = 1 - \widetilde{\gamma}_{0n}^{\text{T}}\gamma_{mn}^{-1} \widetilde{\gamma}_{0n}\,, \quad \widetilde{\gamma}_{0n} = \frac{d \gamma_{0n}}{d \dot{\beta}}\, , \\
			f_2 & = 2 (h_{\beta i} \dot{\zeta}^i +h_{\beta \bar{i}}\dot{\bar{\zeta}}^i)- \left(\widetilde{\gamma}^{\text{T}}_{0n} \gamma_{mn}^{-1}\hat{\gamma}_{0n}+\hat{\gamma}_{0n}^{\text{T}}\gamma_{mn}^{-1} \widetilde{\gamma}_{0n}\right), \quad \hat{\gamma}_{0n} = \gamma_{0n} - \dot{\beta} \widetilde{\gamma}_{0n} \,,\\
			f_3 &= \gamma_{00}-  2 (h_{\beta i} \dot{\zeta}^i +h_{\beta \bar{i}}\dot{\bar{\zeta}}^i)- \hat{\gamma}_{0n}^{\text{T}} \gamma_{mn}^{-1} \hat{\gamma}_{0n}\,.
		\end{split}
	\end{align}
	The specific embeddings that we study are such that $f_{1,2} \rightarrow 0$  as we approach the point where $z_I=0$ ($K \rightarrow \infty$) and, in particular $f_1 \sim e^{- \Delta K}$ , which drastically simplifies the analysis of fluctuations.

		\subsection{Fluctuation analysis near a generic maximal giant}
	
	The maximal giants consist of D3-branes wrapping topologically trivial unions of topologically non-trivial 3-cycles. We choose local complex coordinates on the transverse base $B_4$ such that when the condition $z_I =0$ defining a supersymmetric cycle is satisfied, at least one complex coordinate in $B_4$ tends to infinity. Clearly, this suggests that we are approaching a region where the coordinate patch we are using is not well-defined anymore, but the fluctuation analysis will precisely take place near this edge. In particular, we can choose a gauge for the transverse K\"ahler potential $K(|\zeta^1|, \, |\zeta^2|)$ such that it diverges near points where $z_I=0$ ($\zeta^i\rightarrow \infty$). From the case of the sphere, we learned that $f_1\sim {\rm e}^{-\Delta K}$ appeared in front of $\dot{\beta}^2$ in the DBI action. 
Since we are interested in the maximal giant configurations, a natural coordinate to use is $ \delta f_1=\delta({\rm e}^{-  \Delta K}), \,\, \text{with}\,\, \Delta\in \mathbb{R}_{>0}$, which vanishes at the maximal giants.

For the Wess-Zumino part of the Lagrangian we recall that $C_4$ can be chosen such that:
\begin{align}
	C_4 & = \frac{1}{2}  \mathcal{A}  \wedge J \wedge d \beta\,,
\end{align}
where we choose a gauge in which $C_4$ is regular near the points where the maximal giant configuration is located, in particular
\begin{align}
	\begin{split}
		C_4 & = \mathcal{G}(|\zeta^1|, |\zeta^2|)
		\left(\zeta^1 d \bar{\zeta}^1 - \bar{\zeta}^1 d \zeta^1\right)\wedge d \zeta^2 \wedge d \bar{\zeta}^2 \wedge d \beta \\
		\mathcal{L}_{\text{WZ}}	&  = P[C_4] = \mathcal{G}(|\zeta^1|, |\zeta^2|)
		\epsilon^{0 m n l}\left( \frac{\partial \zeta^2}{\partial \sigma^m}\frac{\partial \bar{\zeta}^2}{\partial \sigma^n}\right) \left( \frac{\partial \bar{\zeta}^1}{\partial \sigma^l}\zeta^1 - \frac{\partial \zeta^1}{\partial \sigma^l} \bar{\zeta}^1 \right) \dot{\beta} \,, 
	\end{split}
\end{align}
where $\mathcal{G}(|\zeta^1|,\, |\zeta^2|)$ is a function that ensures constant behavior as $|\zeta^i|\rightarrow \infty$.
Using our chosen embeddings \eqref{eq:embeddingeqs} and the fact that the contraction with the Levi-Civita symbol in the definition of the pullback ensures that the only place a $\sigma^0$ derivative appears is on the $\beta$ term, we see that we have an overall  $\dot{\beta}$ and no other $\sigma^0$ derivative elsewhere. 
Therefore, in the limit of interest to us, namely close to the maximal giant configurations, the behavior of $\mathcal{L}_{\text{WZ}}$ is such that it is linear in $\dot{\beta}$ and we arrive at the generic form of the Wess-Zumino action
\begin{align} \label{eq:deltaWZ}
	\mathcal{L}_{WZ}& = A_1  + A_2 \dot{\beta}\,, 
\end{align}
where $A_{1,2}$ will be constant values depending on the gauge choice and the specific SE$_5$ analyzed. 
We parameterize the fluctuations as follows:
\begin{align}
	\begin{split}
		\beta(\sigma^0) &= \frac{\omega}{L} \sigma^0+\delta \beta(\sigma^0) \,, \\
		f_1(\sigma^0, \, \sigma^3)&= f_1^{^{(0)}} + \delta f_1(\sigma^0\, , \sigma^3)  =   f_{1,i} \delta \zeta^i + f_{1, \bar{i}} \delta \bar{\zeta}^{\bar{i} } \,,\\
		f_{1,0}  & = f^{(0)}_{1,0} + \frac{\partial \delta f_1}{\partial \sigma^0}\,, \\
			f_{1,n}& = f_{1,n}^{(0)} + \frac{\partial \delta f_1}{\partial \sigma^n}
		\,. 
	\end{split}
\end{align}
Here the coordinates $\delta f_1(t)$ and $\delta \beta (t)$ will play the role of $\delta \xi(t)$ and $\delta \psi(t)$ in the previous case of $S^5$. Then the fluctuations will be parametrized by two variables $\delta f_1$ associated with the radial direction and $\delta \beta(t)$ will be the phase.
The term \eqref{eq:deltaWZ} will produce the action of the perpendicular magnetic field in the effective Lagrangian describing the dynamics of fluctuations about the maximal giants. When analyzing the specific example of $T^{1,1}$, we will see that there is a conical deficit in the space parametrizing the fluctuations.
It is worth highlighting here that in this section we have directly exploited the complex structure on the K\"ahler base.  This structure can be seen to naturally descend from the complex structure of the cone over SE$_5$ which is a local Calabi-Yau space. These are the natural coordinates that figure in the approach to giant gravitons introduced by Mikhailov \cite{Mikhailov:2000ya} and have been shown to be quite natural in the study of more general classical brane configurations in AdS$_5\times$ SE$_5$ spacetimes \cite{Arean:2004mm,Canoura2006,Canoura:2006es}. The element that is essentially new in our work is going beyond the classical configuration and investigating quantum fluctuations in the framework.

\section{The giant graviton expansion in AdS$_5\times T^{1,1}$ } \label{sec:ExampleT11}
In this section, following the general framework discussed in the previous section, we study giant graviton fluctuations in AdS$_5\times T^{1,1}$ and analyze their contribution to the protected superconformal index from the gravity side. We first quantize separately fluctuations about individual cycles $z_I=0$ (in particular, $z_1$ and $z_3$). We show how these problems provide elementary building blocks for more involved configurations when studied using appropriate generalizations of the Quantum Hall matrix model \eqref{eq:QHMM0}.

\subsection{Formulation of the problem}
Let us now consider the background:

\begin{align}\label{eq:T11metric}
\begin{split}
	ds^2&= L^{2} ds^2_{AdS_5}+L^{2} ds^2_{T^{1,1}}, \\
	ds^2_{AdS_5}&= -(1+\frac{r^2}{L^2})dt^2 + \frac{dr^2}{1+\frac{r^2}{L^2}} + r^2 d\Omega_3^2, \\
	ds^2_{T^{1,1}}&= \frac{1}{9}\left(d\psi+\cos\theta_1 d\phi_1+\cos\theta_2 d\phi_2\right)^2 +\frac{1}{6}(d\theta_1^2 +\sin^2\theta_1 d\phi_1^2)+\frac{1}{6}(d\theta_2^2 +\sin^2\theta_2 d\phi_2^2)  \\ 
	F_5&= {\cal F}+ \ast {\cal F},  \\
	{\cal F}&=\frac{4}{L} {\rm vol}(L^5 ~T^{1,1})=\frac{L^4}{27}d\psi\wedge \sin\theta_1\sin\theta_2 d\theta_1\wedge d\theta_2\wedge d\phi_1\wedge d\phi_2 = d\psi \wedge {\rm vol}(\mathbb{P}^1\times \mathbb{P}^1)\,. 
\end{split}
\end{align}
To connect with the metric given in \eqref{eq:ThemetricYpq} we need to set $c=0$ there and identify $\theta_1=\theta, \, \phi=\phi_1, \, y= \cos \theta_2, \, \alpha= - \phi_2$. In this way, we have $Y^{1,0}=T^{1,1}$, which is the base of the conifold. Using the specialization of \eqref{eq:GLSM} for $p=1, \, q=0$. We obtain:
\begin{align} \label{eq:GLSMT11}
|z_1|^2 +|z_2|^2- |z_3|^2-|z_4|^2 & =0\, .
\end{align}
The gauge invariant variables $w_J$ are:
\begin{align}
w_1 & = z_1 z_3, \, \quad w_2 = z_2 z_4\, ,\quad w_3=z_1 z_4, \, \quad w_4 = z_2 z_3\,,
\end{align}
satisfying the constraint $w_1 w_2 = w_3 w_4 $. The angular variables used in \eqref{eq:T11metric}  are related to the complex ones as follows:
\begin{align}\label{eq:zICone}
\begin{split}
	z_1 &= \sin\frac{\theta_1}{2}\, e^{\frac{i}{2}(\psi-\phi_1)}, \quad 
	z_2 = \cos\frac{\theta_1}{2}\, e^{\frac{i}{2}(\psi+\phi_1)}, \\
	z_3 &= \sin\frac{\theta_2}{2}\, e^{-\frac{i}{2}\phi_2}, \quad \quad 
	z_4 = \cos\frac{\theta_2}{2}\, e^{\frac{i}{2}\phi_2}\, ,
\end{split}
\end{align}
in terms of which we can write
\begin{align} \label{eq:zCY}
\begin{split}
	w_1 & = \sin \frac{\theta_1}{2} \sin \frac{\theta_2}{2} {\rm e}^{\frac{i}{2}(\psi - \phi_1 - \phi_2)}, \, \quad
	w_2  = \cos \frac{\theta_1}{2} \cos \frac{\theta_2}{2} {\rm e}^{\frac{i}{2}(\psi + \phi_1 + \phi_2)}, \, \\
	w_3 & =  \cos \frac{\theta_1}{2} \sin \frac{\theta_2}{2} {\rm e}^{\frac{i}{2}(\psi + \phi_1 - \phi_2)}, \, \quad
	w_4  =  \sin \frac{\theta_1}{2} \cos \frac{\theta_2}{2} {\rm e}^{\frac{i}{2}(\psi - \phi_1 + \phi_2)}. \, \\
\end{split}
\end{align}


\subsection{Branes wrapping supersymmetric cycles}
Consider a linear polynomial in $z_1$ defining a holomorphic surface on the conifold
\begin{align}
z_1 = b\,.
\end{align} 
Then the rotation about the Reeb direction $\beta = \frac{\omega t}{L}$ is implemented by making 
\begin{align} \label{eq:constraintZ1}
{\rm e}^{- i \frac{\omega t}{L}}\sin\frac{\theta_1}{2}\, e^{\frac{i}{2}(\psi-\phi_1)} -  b =0\ .
\end{align}
Notice that, since the D3-brane wraps a topologically non-trivial 3-cycle, the rotation of the phase along the Reeb direction is unnecessary, as all the angular momentum is carried by the five-form flux. Only for 3-cycles that become trivial, the rotation parameter has to be the speed of light \cite{Mikhailov:2000ya}, $\omega=1$. We keep $\omega$ generic as a bookkeeping device analogous to the case of $S^5$.
Then, recalling the relation
\begin{align}\label{eq:ReebRotation}
\beta = \frac{1}{3}(\psi - \phi_1 - \phi_2) =\frac{\omega}{L} t\,,
\end{align}
we have
\begin{align}
{\rm e}^{ i \left( \phi_2+ \frac{\omega t}{L}\right)}\sin\frac{\theta_1}{2}\,  - b =0\,.
\end{align}
Consider the supersymmetric cycle defined by 
\begin{align}
z_1 =\sin\frac{\theta_1}{2}\, e^{\frac{i}{2}(\psi-\phi_1)}=0\,.
\end{align}
This reduces \eqref{eq:GLSMT11} to
\begin{align}\label{eq:deltaGLSMT11}
|z_2|^2 - |z_3|^2-|z_4|^2 & =0\,.
\end{align}
We can turn on fluctuations $|\delta z_1|^2$ and turn on small values of a Fayet-Iliopoulos parameter on the right hand side of \eqref{eq:deltaGLSMT11}, such that:
\begin{align} \label{eq:deltaFI1}
\delta z_1(t)^2+	|z_2|^2 - |z_3|^2-|z_4|^2 & =\delta b_1(t)^2\,,
\end{align}  
such that
\begin{align} \label{eq:constB1}
\sin\frac{\delta\theta_1}{2}^2\,  - \delta b_1(t)^2=0\,.
\end{align}
This has a nice interpretation, since small fluctuations of the Fayet-Iliopoulos parameter will play the role of edge modes in the Quantum Hall effect picture. It is illuminating to turn on the fluctuations at the level of \eqref{eq:deltaFI1}, because in this way the constraint \eqref{eq:constB1} depends on the $U(1)$ charge $B_I$ characterizing the supersymmetric cycle that the D3-brane wraps. 
Now we can discuss the problem in the language of section \ref{sec:GeneralKE}. \par
For $T^{1,1}$, one has that the transverse K\"ahler base is $\mathbb{P}^1\times \mathbb{P}^1$ and its K\"ahler potential can be written as:
\begin{equation} \label{eq:Kpotential}
K(\zeta^i, \bar{\zeta}^i)=\frac{2}{3}\sum\limits_{i=1}^2 \ln\left(1 +\frac{3}{2}\zeta^i\bar{\zeta}^i\right).
\end{equation}
We now proceed to write the change of variables from the local complex coordinates to the coordinates in \eqref{eq:T11metric}.
We will choose local complex coordinates for each $\mathbb{P}^1$ 
such that the maximal giant configuration is located at infinity in at least one of the two coordinates to maintain the property we observed in the case of $S^5$. 
\begin{align} \label{eq:P1metric}
\begin{split}
	\zeta^1 & = \sqrt{\frac{2}{3}} \cot \frac{\theta_1}{2} {\rm e}^{~ -i \phi_1}=\sqrt{\frac{3}{2}}\frac{\bar{w}^{2}}{\bar{w}^{4}} = \sqrt{\frac{3}{2}}\frac{\bar{w}^{3}}{\bar{w}^{1}},  \\
	\zeta^2 &  = \sqrt{\frac{2}{3}} \cot \frac{\theta_2}{2} {\rm e}^{~- i \phi_2} =  \sqrt{\frac{3}{2}} \frac{\bar{w}^{2}}{\bar{w}^{3}}=\sqrt{\frac{3}{2}} \frac{\bar{w}^{4}}{\bar{w}^{1}}  \,.
\end{split}
\end{align}
The metric of each $\mathbb{P}^1$ is given by 
\begin{eqnarray}
ds_{i}^2 = \frac{d \zeta^i d\bar \zeta^i}{\left(1+\frac{3}{2}\zeta^i \bar \zeta^i\right)^2}\,.
\end{eqnarray}
Using the definition of $\beta = (\psi - \phi_1 - \phi_2)/3$ from \eqref{eq:ReebRotation}, we can rewrite the metric \eqref{eq:SEmetricK} in the form presented in \eqref{eq:T11metric}. The constraint \eqref{eq:constraintZ1} yields:
\begin{align} \label{eq:EmbHol}
\begin{split}
	{\rm e}^{- \frac{3 K}{2}} &= \sin^2 \frac{\theta_1(\sigma_0, \, \sigma_3)}{2} \sin^2 \frac{\theta_2(\sigma_0, \, \sigma_3)}{2} = \left(1 - \chi^2\right)\sin^2 \frac{\theta_2(\sigma_0, \, \sigma_3)}{2}\\
	\sigma_0 & =t , \, \sigma_{1,2} = \phi_{1,2}, \quad \beta = \frac{\omega}{L} t
\end{split}
\end{align}

The D3-brane wraps the supersymmetric cycle by setting $\theta_1 =0$ and $0<\theta_2<\pi$.
\begin{align}
{\cal F}&=4L^4 {\rm vol}(T^{1,1})=\frac{L^4}{27}d\psi\wedge \sin\theta_1\sin\theta_2 d\theta_1\wedge d\theta_2\wedge d\phi_1\wedge d\phi_2 \,,  \\ 
&=\frac{8 L^4\,}{3\, \zeta_1 \bar{\zeta}_1 \,(3\, \zeta_1 \bar{\zeta}_1 + 2)\,(3\, \zeta_2 \bar{\zeta}_2 + 2)^2} 
\big( \zeta_1\, d\bar{\zeta}_1 - \bar{\zeta}_1\, d\zeta_1 \big) \wedge d \zeta_2 \wedge d \bar{\zeta}_2 \wedge d \beta\,. 
\end{align}


Let us choose the following simplified version of the embedding of the complex coordinates of the transverse $\mathbb{P}^1 \times \mathbb{P}^1$:
\begin{align} \label{eq:EmbedMax}
\begin{split}
	\zeta^1 & = \sqrt{\frac{2}{3}}\cot \frac{\theta_1(\sigma_0)}{2} {\rm e}^{-i \sigma_1} \, ,\\
	\zeta^2 & = \sqrt{\frac{2}{3}}\cot \frac{\theta_2(\sigma_3)}{2} {\rm e}^{-i \sigma_2} \, , 
\end{split}
\end{align}
We are ready to study the dynamics of the fluctuations, for which we set $\beta(t) = \frac{\omega}{L} t -\delta \beta(t)$. Furthermore, near the maximal giant we have that $\theta_1(t) = 0 + \delta \xi_1(t)$. Expanding up to second order in fluctuations will allow us to capture the behavior of fluctuations near the maximal giant. The contributions to the D3 brane Lagrangian are given as:
\begin{align}
\begin{split}
	\mathcal{L}^{(1)}_{\text{DBI}} & = \frac{L^4}{18} \cos\frac{\sigma_3}{2} \cos\frac{\delta \xi_1 }{2}
	\sqrt{
		72  {\rm e}^{- \frac{3}{2}K}\, \dot{\beta}^2
		- \Big((\cos\sigma_3 + 3) \cos\delta \xi_1  + 3 \cos\sigma_3 - 7\Big)\left( \delta \dot{\xi}_1^2 - \frac{6}{L^2}\right)
	}\,, \\
	\mathcal{L}^{(1)}_{\text{WZ}} & = \frac{L^4}{9} \sin \sigma_3\, \dot{\beta} \left(\cos \delta \xi_i-1\right)\, \quad i =1,2 \, \quad {\rm e}^{- \frac{3}{2}K}(\sigma_3, \delta \xi_1) = \sin^2\frac{\sigma_3}{2} \sin^2\frac{\delta \xi_1 }{2}\, .
\end{split}
\end{align}
Note that the functions parameterizing the fluctuations are not independent due to the constraint
\begin{align}\label{eq:deltaConstraint1}
\sin^2 \frac{\delta \xi_1(t)}{2}  & =1- (1- \delta \chi (t))^2\,.
\end{align}

\subsection{The Fock-Darwin system on a cone}
\begin{figure}[h] 
\centering
\includegraphics[width=0.80\textwidth]{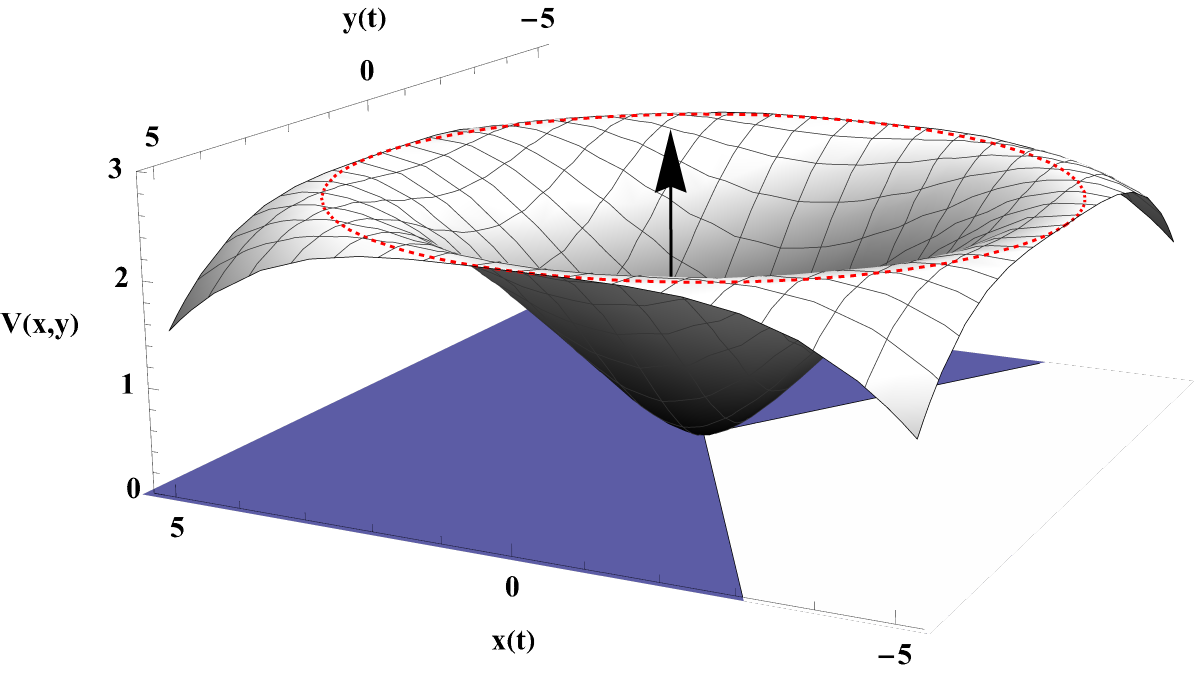}
\caption{The particle moves in a conical background under the action of a constant magnetic field oriented perpendicular to the plane (this is indicated by the vertical arrow in the figure).  The conical deficit is represented by removing a wedge to the blue shaded plane. Additionally, there is an external confining potential. The external potential has a minimum at the center, corresponding to the maximal giant graviton. The maximum is located at $\rho = 4 \sqrt{e N}$, where the fluctuation analysis breaks down. }
\label{fig:PotentialVXy}
\end{figure}

The total action is then given by:
\begin{align}
\mathcal{L}_{\text{D3}} &  = - 4 \pi^2 T_3 \int_{0}^{\pi} d \,\sigma_3 \, \left(\mathcal{L}_{\text{DBI}} - \mathcal{L}_{\text{WZ}} \right)\,,
\end{align}
where we use units such that $4 L^4 = 27 \pi N$ and $T_3 = 1/(2\pi^3)$.
The $\sigma_3$ integral over the WZ term is simple and produces a factor of $2$. Although, the integral of the DBI Lagrangian is nontrivial, it can be performed exactly. Noting that the differential is completed if we set $u= \sin \frac{\sigma_3}{2}$, we have:
\begin{align}
\int_{0}^{\pi} d \, \sigma_3 \,\mathcal{L}_{\text{DBI}} (\sigma_3) = 2 \int_{0}^1 d\, u \,\mathcal{L}_{\text{DBI}}(u)\,.
\end{align}
The exact result is not particularly illuminating, but it drastically simplifies after expanding in terms of small fluctuations up to second order. 
Let us perform the following change of variables: 
\begin{align} \label{eq:changeofvariables}
\delta \xi & \equiv \rho = \sqrt{ \frac{x^2 + y^2}{L^2}}\, \geq0\, , \quad \delta \beta \equiv \varphi = \frac{2}{3} \arctan \frac{y}{x}\,.
\end{align} 
Absorbing a factor of $L$ in the definition of the Lagrangian, we rescale the variables $(x, y, t) \rightarrow \left(\frac{L x}{\sqrt{N}}\,, \frac{L y}{\sqrt{N}}, \frac{t}{L}\right)$ and describe the fluctuations by the following Lagrangian
\begin{align} \label{eq:LanLagrangian}
\begin{split}
	\mathcal{L}_{\text{D3}} &= \frac{1}{2} M \left(\dot{x}^2+\dot{y}^2\right)+ \frac{1}{2}  B \left(x \dot{y}-y \dot{x}\right)- \frac{3}{4}N -V(x,y) \, , \\
	V(x, y)& =  \frac{M}{2 } \left(x^2+y^2\right) \left( \Omega^2- 3\log \left(\frac{1}{4 } \sqrt{\frac{x^2+y^2}{N}}\right)\right) \,, \\
	 \textrm{with} \quad M & = \frac{1}{8 }, \, \,  B =  \frac{1}{8 }(4- 3 \omega), \quad \omega_c \equiv \frac{ B}{M} = 4 - 3 \omega  \,, \\
	\Omega^2 & = \frac{3}{4}(\omega\left(8-3 \omega ) -1\right)=-\frac{1}{4}\left(\omega_c- \sqrt{13}\right)\left(\omega_c+\sqrt{13}\right).
\end{split}
\end{align}
Let us immediately note that this already marks a departure from the $S^5$ case. From the range of $\varphi$ in \eqref{eq:changeofvariables}, we see that the geometry is that of a cone, with a deficit angle given by 
\begin{eqnarray} \label{eq:deficit}
\Delta \varphi = 2\pi -  2\pi \frac{2}{3}  = \frac{2\pi}{3}\,.
\end{eqnarray}
This Lagrangian therefore corresponds to a charged particle of mass $M$ moving on a cone with angular deficit $\frac{2 \pi}{3}$ and under the combined action of a constant magnetic field perpendicular to the surface and an external potential that has the form $V(\rho) \sim \rho^2 \left(a +b \log \rho^2\right)$. 
In figure \ref{fig:PotentialVXy}, we show a plot of the potential with $N=1$, where we see which the absolute maximum is located at $\rho_* =4 \sqrt{ e N}$,  
a region where our fluctuation analysis near the maximal giant should break down. For the quantization process we shall focus on the large $\omega_c$ regime, that ensures that the logarithmic part of the potential becomes parametrically subdominant, that is subdominant with respect to the quadratic potential.

The constant term in  $\mathcal{L}_{\text{D3}} $ produces exactly the contribution of the energy of the maximal giant in the classical configuration, which precisely matches the volume wrapped by the D3-brane $\mathcal{E}_0=8 \pi^2 T_3 L^4 = \frac{3}{4}N$. The Lagrangian associated with the Fock-Darwin system is of the following form:
\begin{align}\label{eq:LFD}
\mathcal{L}_{\text{FD}} & =  \frac{1}{2} M \left(\dot{x}^2+\dot{y}^2\right)+ \frac{1}{2}  B \left(x \dot{y}-y \dot{x}\right) -\frac{M}{2 L} \Omega^2 (x^2 + y^2)- \frac{3}{4}N\, .
\end{align}


\subsection{The $T^{1,1}$ giant graviton index from the Quantum Hall effect.}
Here, we would like to adopt the same language as in section \ref{Sec:ReviewS5} to describe the lowest Landau level Hilbert space in the language used in the Quantum Hall effect.
Following a similar procedure as in section \ref{Sec:ReviewS5}, we can define complex coordinates as in \eqref{eq:zzbar}, but keeping in mind that their phase is not $2 \pi$ periodic due to the angular deficit \eqref{eq:deficit}, we obtain the Lagrangian
\begin{align} \label{eq{LFD}}
	\begin{split}
		\mathcal{L}_{\text{FD}} & =  M \dot{z} \dot{\bar{z}} + \frac{i M \omega_c}{2}\left(z \dot{\bar{z}} - \bar{z} \dot{z}\right)- M\Omega^2 z \bar{z} -  \frac{3}{4}N \,.
	\end{split}
\end{align}
Focusing on the lowest Landau level will be our goal, as we have identified its Hilbert space with the one associated with $\frac{1}{2}$-BPS fluctuations. Taking the $|\omega_c| \rightarrow \infty$ limit of \eqref{eq:LFD}, we obtain
\begin{align}
	\mathcal{L}_{\text{FD-LL}} & = \frac{i M \omega_c}{2}\left(z \dot{\bar{z}}- \bar{z} \dot{z}\right) - M \Omega^2 z \bar{z} - \frac{3}{4}N \,.
\end{align}
We define canonical momenta as
\begin{align}
	p_z & = \frac{\partial \mathcal{L}_{\text{Lan}}
	}{\partial \dot{z}} =  - \frac{i M \omega_c}{2} \bar{z}\, , \quad p_{\bar{z}} =  \frac{\partial \mathcal{L}_{\text{Lan}}
	}{\partial \dot{\bar{z}}} = \frac{i M \omega_c}{2} z\, .
\end{align}
The Hamiltonian of this problem is 
\begin{align}
	\mathcal{H}_{\text{FD-LL}}(\Omega) & =  \Omega^2 z \bar{z} +\frac{3}{4}N\,.
\end{align} From now on, we will set the mass $M=1$. We consider \eqref{eq:deltaConstraint} and expand for small $\delta \xi_1^2 \sim z \bar z ,$:
\begin{align} \label{eq:T11constraint1}
	 z \bar{z}(\sigma_0) & = 1- \chi \bar{\chi}(\sigma_0)\,.
\end{align}
Then we insert a $U(1)$ gauge field and construct the $U(1)$  effective theory:
\begin{align}
	\mathcal{L}_{U(1)_1} & =- i \omega_c z D_t \bar{z} - \Omega^2 z \bar{z} - \frac{3}{4}N
	-i\omega_c \bar{\chi}D_t \chi + \omega_c a_0 \,
\end{align}
such that
\begin{align}
	D_t z &= \dot{z}- i a_0 z\, ,  \quad D_t \chi = \dot{\chi} - i a_0 \chi\, .
\end{align}
The equations of motion of $a_{0}$ yield \eqref{eq:T11constraint1}. We are now in a position to promote our problem to its non-Abelian version in order to capture multiple giant gravitons.  

\begin{align}
	\begin{split}
		\mathcal{L}_{U(m)_1} &= 
		-i\omega_c \, \mathrm{Tr} \Big( 
		Z D_t Z^\dagger + \chi^\dagger D_t \chi
		\Big)  -\Omega^2 \, \mathrm{Tr} \Big( Z^\dagger Z  \Big)
		+ \omega_c \, \mathrm{Tr}  A_0  - \frac{3}{4}m N
		,\\
		D_t Z &= \dot Z - i [A_0, Z]\,, \quad 
		D_t \chi = \dot \chi - i A_0 \chi \, ,
		\quad [Z,Z^\dagger] + \Phi \Phi^\dagger =   \mathbbm{1}_m \, .
	\end{split}
\end{align}
We define the partition function of this system as
\begin{align}
	\mathcal{Z}(q, x) & =q^{\frac{3}{4}m N}\text{Tr} \,{\rm e}^{- \gamma ( \frac{\Omega^2}{|\omega_c|} n- \mu j)} , \quad q \equiv {\rm e}^{- \gamma \frac{\Omega^2}{|\omega_c|}}, \quad x \equiv {\rm e}^{\gamma \mu}\, ,
\end{align}
where we have included the vacuum energy. Using the result of Appendix \ref{App:QHE}, specifically \eqref{eq:ZQHE} for $k=1$ we obtain:
\begin{align}
	\mathcal{Z}(q, \, x) & =  x^{ m} {\rm e}^{- \gamma\frac{3}{4}m N}
	q^{ \frac{m (m-1)}{2}} \prod_{j=1}^m \frac{1}{(1- q^j)} 
\end{align}
Similarly to the $^5$ case, an index can be obtained if we shift the chemical potential by $\frac{i \pi}{\gamma }$, hence $x^j \rightarrow {\rm e}^{\gamma\mu  (j+ \frac{i \pi}{\gamma \mu}j)}=(-1)^j x^{j}$. We define:
\begin{align}\label{eq:ImT11x1}
	\begin{split}
		\mathcal{I}_{U(m)_1}(q, x) 
		& =  (-1)^mx^{ m} {\rm e}^{- \gamma\frac{3}{4}m N}q^{ \frac{ m (m-1)}{2}} \prod_{j=1}^m \frac{1}{(1- q^j)}\,. 
	\end{split}
\end{align}
In order to impose a relation between $x$ and $q$ in a way that the index is graded by $H-R$, we notice that the charges of the fluctuations on the brane and those on the AdS$_5 \times T^{1,1}$ are
\begin{align} \label{eq:R23}
	i \frac{2}{3}\partial_{\varphi} & = -i \partial_{\beta}  =  R\,, \quad i \frac{\Omega^2}{|\omega_c|} \partial_t - i \partial_{\varphi} = \frac{\Omega^2}{|\omega_c|} H - \frac{3}{2} R\,,
\end{align}
which means that the lowest Landau level Hilbert space can be mapped to that of fluctuations of the BPS branes $H-R$, if $\frac{\Omega^2}{|\omega_c|} = \frac{3}{2}$ which again yields the natural regularization $q ={\rm e}^{- \frac{3}{2}\gamma}$. This is consistent with the analysis below \eqref{eq:Iqtilde}, where we can see that $\widetilde{q}^4 = q^2$. To match the vacuum contribution, we fix the normalization such that $\widetilde{q} = {\rm e}^{- \frac{3}{4}\gamma}$, which in turn yields $q = {\rm e}^{-\frac{3}{2} \gamma}$. We emphasize that this choice reflects a regularization ambiguity inherent in the effective matrix model description of the Landau/Fock-Darwin system, rather than an ambiguity of the underlying protected index. 

From this perspective, different choices of $q$ correspond to different normalizations of the effective quantum mechanical Hamiltonian, while the physical input is fixed by the requirement of matching the lowest Landau level contribution to the local building blocks entering the protected sector. In this sense, the regularization is not unique at the level of the effective model, but it is uniquely fixed once the matching to the semiclassical giant graviton sector is imposed.
Implementing this transformation in \eqref{eq:ImT11x1} yields
\begin{align}
	\begin{split}
		\mathcal{I}_{U(m)_1}(q)& = (-1)^m q^{\frac{1}{2}m N }q^{ \frac{m (m+1)}{2}} \prod_{j=1}^m \frac{1}{(1- q^j)} \\
		&=(-1)^m {\rm e}^{-  \frac{3}{4} \gamma m N}{\rm e}^{-\gamma\frac{3 m (m+1)}{4}} \prod_{j=1}^m \frac{1}{(1- {\rm e}^{- \frac{3}{2} \gamma j})}\,.
	\end{split}
\end{align}


\subsection{Quantum Hall effect with  $U(m)\times U(m)$ and bifundamental edge modes } \label{sec:Bifundamental}
\begin{figure}[h!]
\centering
\begin{tikzpicture}[
tdplot_main_coords,
font=\footnotesize,
Helpcircle/.style={gray!70!black, thin},
scale=1.2
]
\begin{scope}[shift={(0,0,0)}] 

\def\circleamp{1.2*\rlat/2} 
\def\arrowfrac{0.9} 

\begin{scope}[shift={(-\r-\gap,0,0)}]
  \fill[ball color=blue!90, opacity=0.4] circle (\r);

  \fill[orange!90, opacity=0.6]
    plot[domain=0:360, samples=200, variable=\t]
      ({4/5*cos(\t)}, {4/5*sin(\t)}, {\h}) -- (0,0,\h) -- cycle;
 \fill[white!90]
    plot[domain=0:360, samples=200, variable=\t]
      ({4/10*cos(\t)}, {4/10*sin(\t)}, {\h}) -- (0,0,\h) -- cycle;

  \draw[thick, ->, black, >=stealth] 
    plot[domain=30:300, samples=100, variable=\t]
      ({\arrowfrac*\circleamp*cos(\t)}, {\arrowfrac*\circleamp*sin(\t)}, {\h});

  \coordinate (M1b) at (0,0,\h);
  \shade[ball color=black] (M1b) circle (0.75pt);
  \node at (M1b) [yshift=32pt] {$\theta_{\text{min}}<\theta_1<\theta_{\text{max}}$};
  \node at (M1b) [xshift=21pt] {$\phi_1$};
\end{scope}

\begin{scope}[shift={(\r+\gap,0,0)}]
  \fill[ball color=orange!90, opacity=0.6] circle (\r);

  \fill[blue!90, opacity=0.5]
    plot[domain=0:360, samples=200, variable=\t]
      ({4/5*cos(\t)}, {4/5*sin(\t)}, {\h}) -- (0,0,\h) -- cycle;

 \fill[white!90]
    plot[domain=0:360, samples=200, variable=\t]
      ({4/10*cos(\t)}, {4/10*sin(\t)}, {\h}) -- (0,0,\h) -- cycle;

  \draw[thick, ->, black, >=stealth] 
    plot[domain=30:300, samples=100, variable=\t]
      ({\arrowfrac*\circleamp*cos(\t)}, {\arrowfrac*\circleamp*sin(\t)}, {\h});

  \coordinate (M2b) at (0,0,\h);
  \shade[ball color=black] (M2b) circle (0.75pt);
  \node at (M2b) [yshift=32pt] {$\theta_{\text{min}}<\theta_2 < \theta_{\text{max}}$};
  \node at (M2b) [xshift=21pt] {$\phi_2$};
\end{scope}

\draw[thick,->] (0,-1) -- (0,1.5) node[anchor=south]{$\beta = \frac{\omega}{L} t$}; 
\draw[-latex, thick, rotate=-90] (0,0.1) arc [start angle=-190, end angle=153, x radius=0.3cm, y radius=0.6cm];

\end{scope}
 
\end{tikzpicture}
 \caption{A generic giant graviton configuration is shown schematically. We represent the two $S^2$ on the base of the $T^{1,1}$ manifold. The giant wraps two complementary sectors of the spheres separately which are mapped into each other by the constraint \eqref{eq:MikhailovConstraint}. The white region is not part of the worldvolume. As discussed in \cite{Hamilton:2010sv},the azimuthal angles $\theta_{1,2}$ take values in two types of intervals, namely $[\theta_{\text{min}}, \, \theta_{\text{max}}] = [2 \arccos \chi, \, 2 \arcsin (1- \chi^2)^{1/4}]$  and $[\theta_{\text{max}},\, \pi]$. }
\label{fig:GenericGiant}
 \end{figure}
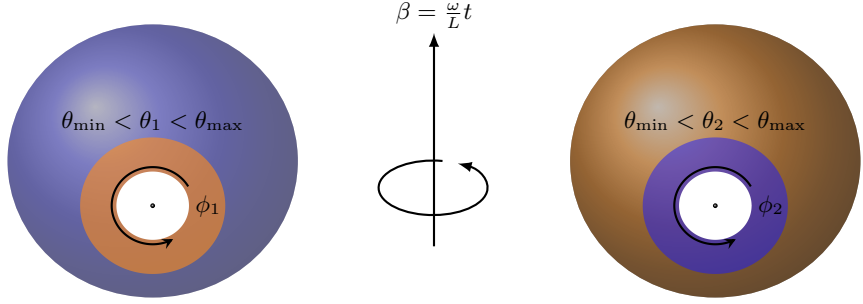

We focus now on the D3-brane configurations that fall within the same category as the original giant gravitons, and on the fluctuation analysis about their maximal configuration.
If we are to implement Mikhailov's embedding defining the giant graviton \cite{Mikhailov:2000ya} (see also \cite{Beasley:2002xv, Berenstein:2002ke}), the $\frac{1}{2}$-BPS configurations are described as holomorphic surfaces depending on only one of the $w_{J=1,\cdots, 4}$ variables. In particular, as analyzed in \cite{Hamilton:2010sv}, we can focus on the configurations described by 
\begin{align} \label{eq:MikhailovConstraint}
   w_1 & = \sqrt{1 - \chi^2}   \, ,
\end{align} 
for $\chi$ real and taking values between $0$ and $1$. The rotation occurs along the Reeb direction, which can be seen from \eqref{eq:zCY}, is given by
\begin{align} \label{eq:ReebRotation0}
    w^1 & = \sin \frac{\theta_1}{2} \sin \frac{\theta_2}{2} {\rm e}^{i \frac{3}{2} \beta(t)}, \quad \beta = \frac{1}{3}(\psi - \phi_1 - \phi_2) =\frac{\omega}{L} t\, .
\end{align}
 It will be instructive to keep this parameter unspecified in some of the manipulations below. In figure \ref{fig:GenericGiant} we sketch the configuration specified by the constraint \eqref{eq:MikhailovConstraint}.
 Now we can discuss the problem in the language of section \ref{sec:GeneralKE}.

 The solution of the maximal giant corresponds to $\chi
 =1$, which can be achieved either by setting $\theta_2=0$ and $0<\theta_1<\pi$ or by setting $\theta_1 =0$ and $0<\theta_2<\pi$.
\begin{align}
 {\cal F}&=4L^4 {\rm vol}(T^{1,1})=\frac{L^4}{27}d\psi\wedge \sin\theta_1\sin\theta_2 d\theta_1\wedge d\theta_2\wedge d\phi_1\wedge d\phi_2 \,,  \\ 
&=\frac{8 L^4\,}{3\, \zeta_1 \bar{\zeta}_1 \,(3\, \zeta_1 \bar{\zeta}_1 + 2)\,(3\, \zeta_2 \bar{\zeta}_2 + 2)^2} 
\big( \zeta_1\, d\bar{\zeta}_1 - \bar{\zeta}_1\, d\zeta_1 \big) \wedge d \zeta_2 \wedge d \bar{\zeta}_2 \wedge d \beta\,. 
\end{align}


Let us choose the following simplified version of the embedding of the complex coordinates of the transverse $\mathbb{P}^1 \times \mathbb{P}^1$:
\begin{align} \label{eq:EmbedMax}
\begin{split}
    \zeta^1 & = \sqrt{\frac{2}{3}}\cot \frac{\theta_1(\sigma_0)}{2} {\rm e}^{-i \sigma_1} \, ,\\
    \zeta^2 & = \sqrt{\frac{2}{3}}\cot \frac{\theta_2(\sigma_3)}{2} {\rm e}^{-i \sigma_2} \, , 
    \end{split}
\end{align}



\begin{figure}[h!]
\centering
\begin{tikzpicture}[
tdplot_main_coords,
font=\footnotesize,
Helpcircle/.style={gray!70!black, thin},
scale=1.2
]

\begin{scope}[shift={(0,0,0)}] 

\def\wiggleamp{0.035}  
\def\wigglefreq{8}     

\begin{scope}[shift={(-\r-\gap,0,0)}]
  \fill[ball color=blue!90, opacity=0.4] circle (\r);

  \fill[orange!90, opacity=0.6] 
    plot[domain=0:360, smooth, samples=400, variable=\t] 
      ({(\rlat/2 + \wiggleamp*sin(\wigglefreq*\t))*cos(\t)}, 
       {(\rlat/2 + \wiggleamp*sin(\wigglefreq*\t))*sin(\t)}, 
       {\h}) -- (0,0,\h) -- cycle;
        \fill[white!90] 
    plot[domain=0:360, smooth, samples=400, variable=\t] 
      ({(\rlat/4 + \wiggleamp*sin(3/4*\wigglefreq*\t))*cos(\t)}, 
       {(\rlat/4 + \wiggleamp*sin(3/4*\wigglefreq*\t))*sin(\t)}, 
       {\h}) -- (0,0,\h) -- cycle;
  \draw[thick, ->, black, >=stealth] 
    plot[domain=30:300, samples=100, variable=\t]
    ({(1.1*\rlat/2 + \wiggleamp)*cos(\t)}, 
     {(1.1
    *\rlat/2 + \wiggleamp)*sin(\t)}, 
     {\h});
  \draw[Helpcircle, domain=0:360, smooth, samples=400, variable=\t]
    plot ({(\rlat/2 + \wiggleamp*sin(\wigglefreq*\t))*cos(\t)}, 
          {(\rlat/2 + \wiggleamp*sin(\wigglefreq*\t))*sin(\t)}, 
          {\h});

  \coordinate (M1b) at (0,0,\h);
  \shade[ball color=black] (M1b) circle (0.75pt);
  \node at (M1b) [yshift=20pt] {$\theta_1 = 0 + \delta \xi_1(t)$};
   \node at (M1b) [xshift=-23pt] {$\phi_1$};
\end{scope}

\begin{scope}[shift={(\r+\gap,0,0)}]
  \fill[ball color=orange!90, opacity=0.6] circle (\r);

  \fill[blue!90, opacity=0.5] 
    plot[domain=0:360, smooth, samples=400, variable=\t] 
      ({(\rlat/2 + \wiggleamp*sin(\wigglefreq*\t))*cos(\t)}, 
       {(\rlat/2 + \wiggleamp*sin(\wigglefreq*\t))*sin(\t)}, 
       {\h}) -- (0,0,\h) -- cycle;

  \fill[white!90] 
    plot[domain=0:360, smooth, samples=400, variable=\t] 
      ({(\rlat/4 + \wiggleamp*sin(3/4*\wigglefreq*\t))*cos(\t)}, 
       {(\rlat/4 + \wiggleamp*sin(3/4*\wigglefreq*\t))*sin(\t)}, 
       {\h}) -- (0,0,\h) -- cycle;
  \draw[Helpcircle, domain=0:360, smooth, samples=400, variable=\t]
    plot ({(\rlat/2 + \wiggleamp*sin(\wigglefreq*\t))*cos(\t)}, 
          {(\rlat/2 + \wiggleamp*sin(\wigglefreq*\t))*sin(\t)}, 
          {\h});
\draw[thick, ->, black, >=stealth] 
    plot[domain=30:300, samples=100, variable=\t]
    ({(1.1*\rlat/2 + \wiggleamp)*cos(\t)}, 
     {(1.1*\rlat/2 + \wiggleamp)*sin(\t)}, 
     {\h});
  \coordinate (M2b) at (0,0,\h);
  \shade[ball color=black] (M2b) circle (0.75pt);
  \node at (M2b) [yshift=20pt] {$\theta_2 = 0 + \delta \xi_2(t)$};
     \node at (M2b) [xshift=-23pt] {$\phi_2$};
\end{scope}

\draw[thick,->] (0,-1) -- (0,1.5) node[anchor=south]{$\beta = \frac{\omega}{L} t -\delta \beta(t)$}; 
\draw[-latex, thick, rotate=-90] (0,0.1) arc [start angle=-190, end angle=153, x radius=0.3cm, y radius=0.6cm];

\end{scope}

\begin{scope}[shift={(0,0,-4*\r - \h - 2)}] 

  \begin{scope}[shift={(-\r-\gap,0,0)}]
    \fill[ball color=blue!90, opacity=0.4] circle (\r);
    \coordinate (M1) at (0,0,\h);
    \shade[ball color=black] (M1) circle (0.75pt);
    \node at (M1) [yshift=20pt] {$\theta_1 = 0$};
    \node at (M1) [xshift=-23pt] {$\phi_1$};
  \draw[thick,->, black, >=stealth] 
      plot[domain=30:300, samples=100, variable=\t]
      ({(1.1*\rlat/2)*cos(\t)}, 
       {(1.1*\rlat/2)*sin(\t)}, 
       {\h});
    
  \end{scope}

  \begin{scope}[shift={(\r+\gap,0,0)}]
    \fill[ball color=orange!90, opacity=0.6] circle (\r);
    \coordinate (M2) at (0,0,\h);
    \shade[ball color=black] (M2) circle (0.75pt);
    \node at (M2) [yshift=20pt] {$\theta_2 = 0$};
     \node at (M2) [xshift=-23pt] {$\phi_2$};
       \draw[thick,->, black, >=stealth] 
      plot[domain=30:300, samples=100, variable=\t]
      ({(1.1*\rlat/2)*cos(\t)}, 
       {(1.1*\rlat/2)*sin(\t)}, 
       {\h});
  \end{scope}

  \draw[thick,->] (0,-1) -- (0,1.5) node[anchor=south]{$\beta = \frac{\omega}{L} t$}; 
  \draw[-latex, thick, rotate=-90] (0,0.1) arc [start angle=-190, end angle=153, x radius=0.3cm, y radius=0.6cm];

\end{scope}

\end{tikzpicture}

 \caption{The maximal giant configuration is shown in the upper part. We represent the two $S^2$ on the base of the $T^{1,1}$ manifold that are separately wrapped by the maximal giant and then mapped into each other by the constraint \eqref{eq:MikhailovConstraint}. Turning on small fluctuations about this configuration, we obtain what is schematically shown in the lower part of the figure. A thin strip is generated whose with is controlled by the deviation of the parameter $\alpha$ from $1$. Note that the points where the origin in the $\delta \xi_{1,2}$ planes ( namely $\theta_1=0(\theta_2=0)$) are always avoided by the system.}
     \label{fig:MaximalGiantFluctuations}
 \end{figure}
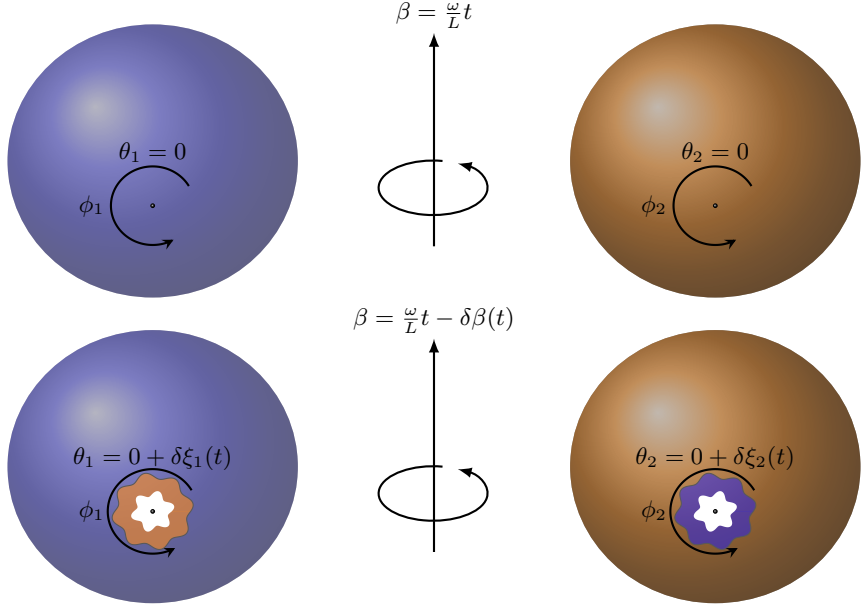

 Before performing the expansion we observe that the proposed embedding \eqref{eq:EmbedMax} violates the holomorphic constraint \eqref{eq:EmbHol} as can be seen if we combine the definition of $w_1$ \eqref{eq:zCY} and \eqref{eq:EmbedMax}:
 \begin{align}
     \sin \frac{\theta_1(\sigma_0)}{2} \sin \frac{\theta_2(\sigma_3)}{2} & \neq \text{constant}\,, \quad \text{unless} \quad \theta_1 =0, \,\,\chi =1\,
 \end{align}
 that corresponds to the maximal giant configuration. Therefore, we need to keep in mind that our embedding \eqref{eq:EmbedMax} will only be valid for performing a fluctuation analysis near the maximal giant configuration. Moreover, there is an equivalent embedding with $\theta_2 =\theta_2(\sigma_0)$ and $\theta_1 = \theta_1(\sigma_3)$, that will only be valid for the maximal giant configuration where $\theta_2 =0$. The two configurations contribute equally to the D3-brane action as we will see below. After having clarified this point, we are ready to study the dynamics of the fluctuations $\delta \beta (t)$ defined via $\beta(t) = \frac{\omega}{L} t -\delta \beta(t)$. Furthermore, near the maximal giant we have that $\theta_1(t) = 0 + \delta \xi_1(t)$ (or $\theta_2= 0 + \delta \xi_2(t)$). We illustrate how this process reflects in the geometry in figure \ref{fig:MaximalGiantFluctuations}. 

 Expanding up to second order in fluctuations will allow us to capture the behavior of fluctuations near the maximal giant. The contributions to the D3 brane Lagrangian are given by
 \begin{align}
 \begin{split}
     \mathcal{L}^{(i)}_{\text{DBI}} & = \frac{L^4}{18} \cos\frac{\sigma_3}{2} \cos\frac{\delta \xi_i }{2}
\sqrt{
72  {\rm e}^{- \frac{3}{2}K}\, \dot{\beta}^2
- \Big((\cos\sigma_3 + 3) \cos\delta \xi_i  + 3 \cos\sigma_3 - 7\Big)\left( \delta \dot{\xi}_i^2 - \frac{6}{L^2}\right)
}\, \\
\mathcal{L}^{(i)}_{\text{WZ}} & = \frac{L^4}{9} \sin \sigma_3\, \dot{\beta} \left(\cos \delta \xi_i-1\right)\, \quad i =1,2 \, \quad {\rm e}^{- \frac{3}{2}K}(\sigma_3, \delta \xi_i) = \sin^2\frac{\sigma_3}{2} \sin^2\frac{\delta \xi_i }{2}\, .
\end{split}
 \end{align}
 The functions parameterizing the fluctuations are not independent due to the constraint:
 \begin{align}\label{eq:deltaConstraint}
     \sin \frac{\delta \xi_1(t)}{2} \sin \frac{\sigma_3}{2} & =  \sin \frac{\delta \xi_2(t)}{2} \sin \frac{\sigma_3}{2}=\sqrt{1- (1- \delta \chi (t))^2}, \, \quad \Rightarrow \, \delta \xi_1 =\delta \xi_2 \equiv \delta \xi \, .
 \end{align}




Recalling that there were two possible choices of embedding yielding a similar Lagrangian and related to each other via \eqref{eq:EmbHol}, we will study:
\begin{align}
\mathcal{L}_{\text{FD-LL}}(\Omega) & =i\frac{M \omega_c}{2}\left(z \dot{\bar{z}}- \bar{z} \dot{z} + w \dot{\bar{w}} - \bar{w}\dot{w}\right) - M \Omega^2 \left( z \bar{z}+w \bar{w}\right) - \frac{3}{4}N\,,\\
   \mathcal{H}_{\text{FD-LL}}(\Omega) & = M \Omega^2(z \bar{z}+ w \bar{w})+\frac{3}{4}N   \,.
\end{align}
 We now need to insert the edge degrees of freedom that implement the constraint between the two coordinates and ensure tha validity of Mikhailov constraint. We consider \eqref{eq:deltaConstraint} and expand for small $\delta \xi_1^2 \sim z \bar z ,\, \delta \xi_2^2 \sim w \bar{w}$:
\begin{align}
  2 z \bar{z}(\sigma_0) \,\sin^2 \frac{\sigma_3}{2} & = 1- \chi \bar{\chi}(\sigma_0, \sigma_3) = 2 w \bar{w}(\sigma_0) \sin^2 \frac{\sigma_3}{2}\,, 
\end{align}
which, after integrating over $\sigma_3 \in [0, 2 \pi]$, becomes
\begin{align} \label{eq:T11constraint}
     z \bar{z}(\sigma_0)  & = 1- \langle\chi \bar{\chi}\rangle=  - w \bar{w}(\sigma_0 ) \, ,
\end{align}
where $\langle \chi \bar{\chi}\rangle(\sigma_0) = \frac{1}{2\pi} \int_{0}^{2 \pi} \chi \bar{\chi} (\sigma_0,\, \sigma_3)d \sigma_3$. The minus sign on the right hand side of \eqref{eq:T11constraint} accounts for the fact that $w\bar{w}$ parametrizes fluctuations about the cycle $z_3=0$, which has negative baryonic charge \eqref{eq:GLSMT11}. 
 Define $\Phi$, such that $\bar{\Phi} \Phi = \langle \bar{\chi} \chi \rangle$, we insert two $U(1)$ gauge fields and construct the $U(1)_1\times U(1)_{-1}$ effective theory:
\begin{align}
\mathcal{L}_{U(1)_1\times U(1)_{-1}} & = -i\omega_c \left(z D_t \bar{z} + w D_t \bar{w}\right) - \Omega^2\left(z \bar{z} + w\bar{w}\right) 
+ i\omega_c \bar{\Phi}D_t \Phi +\omega_c(A_0^{(1)} - A_0^{(2)}) \end{align}
such that
\begin{align}
D_t \bar{z} &= \dot{z}- i A_0^{(1)} z\, , \quad D_t w = \dot{w} + i A_0^{(2)} w\, , \quad D_t \Phi = \dot{\Phi} - i (A_0^{(1)}-A_0^{(2)}) \Phi\, .
\end{align}
The equations of motion of $A_{0}^{(1,2)}$ yield \eqref{eq:T11constraint}. We are now in a position to promote our problem to its non-Abelian version in order to capture multiple giant gravitons.  

\begin{align}
\begin{split}
\mathcal{L}_{U(m)_1\times U(m)_{-1}} &= 
-i \omega_c \, \mathrm{Tr} \Big( 
Z D_t Z^\dagger + W D_t W^{\dagger} + \Phi^\dagger D_t \Phi
\Big)\,, \\
&\quad - \Omega^2 \, \mathrm{Tr} \Big( Z^\dagger Z + W^\dagger W \Big)
+ \omega_c \, \mathrm{Tr} \Big( A_0^{(1)} - A_0^{(2)} \Big) 
\,,\\
D_t Z &= \dot Z - i [A_0^{(1)}, Z]\,, \\
D_t W &= \dot W - i [A_0^{(2)}, W]\,, \\
D_t \Phi &= \dot \Phi - i A_0^{(1)} \Phi + i \Phi A_0^{(2)} \,,\\
 [Z,Z^\dagger] + \Phi \Phi^\dagger - \mathbf{1}  &= 0 \,, \\
 [W,W^\dagger] - \Phi^\dagger \Phi + \mathbf{1}  &= 0 \, .
\end{split}
\end{align}
We define the partition function directly in terms of the regularized fugacity $q ={\rm e}^{- \gamma}$:
\begin{align}
\mathcal{Z}(q, x) & = q^{\frac{3}{2}m N}\text{Tr} \,{\rm e}^{- \gamma ( n- \mu j)} , \quad q \equiv {\rm e}^{- \gamma }, \quad x \equiv {\rm e}^{\gamma \mu}\, ,
\end{align}
where we have included the vacuum energy. To describe the system of $m$ coincident giants, we consider the two gauge group factors $U(m)\times U(m)$. We then generalize the matrix model construction of \cite{Dorey:2016hoj} to the case of two complex matrix-valued oscillators, $Z$ and $W$, transforming in the adjoint of $U(m)_L \times U(m)_R$, with an edge mode type of field $\Phi$ transforming in the bifundamental. Let $z_a$, $a=1,\dots,m$, and $w_b$, $b=1,\dots,m$, denote Cartan fugacities for the two gauge groups. We denote the fugacity for the edge modes as $x$. Ignoring the gauge constraints, the adjoint oscillators contribute
\begin{align}
	\mathcal{Z}_{\rm ZW} = \mathcal{Z}_Z \mathcal{Z}_W= \prod_{a\neq b }^{m} \frac{1}{(1 - q\, z_a/z_b)} \frac{1}{(1 - q\, w_a/w_b)}\,,
\end{align}
while the edge modes contribute
\begin{align}
	\mathcal{Z}_{\Phi} = \prod_{a,b=1}^{m} \frac{1}{1 - \frac{z_a}{w_b} x}\,.
\end{align}
The partition of the full non-gauge invariant Hilbert space is $\mathcal{Z}_{\rm NG} = \mathcal{Z}_{\rm ZW}\,\mathcal{Z}_{\Phi}$.
Projecting onto singlet states via contour integration over the Cartan torus yields
\begin{align}
	\mathcal{Z}(q, x) = \frac{1}{(m!)^2} 
	\prod_{a=1}^{m} \oint_{|z_a|=1} \frac{dz_a}{2\pi i\, z_a^{k+1}} 
	\prod_{b=1}^{m} \oint_{|w_b|=1} \frac{dw_b}{2\pi i\, w_b^{-k+1}} \,
	\Delta(z) \, \Delta(w) \, \mathcal{Z}_{\rm NG}\,,
\end{align}
with Vandermonde determinants $\Delta(z) = \prod_{a \neq a'} (1-z_a/z_{a'})$ and $\Delta(w) = \prod_{b \neq b'} (1-w_b/w_{b'})\,$.
Let us rewrite the edge mode contribution in the following way:
\begin{align}
	\prod_{a, b=1}^m \frac{1}{1- x\frac{z_a}{ w_b}} & =\sum_{\lambda} H_{\lambda}((w/x)^{-1}\,, q)P_{\lambda}(z, \, q) = \sum_{\lambda} H_{\lambda}(w^{-1}\,, q)P_{\lambda}(z x, \, q)\,.
\end{align}
Here we are making use of the Hall-Littlewood polynomials $P_{\lambda}(\zeta\,, q)$, the modified Hall-Littlewood polynomials (or Milne polynomials) $H_{\lambda}(\zeta, \, q)$ and their properties. In Appendix \ref{App:QHE} we summarize the manipulations carried out in \cite{Dorey:2016hoj} as well as the definitions and some properties of these functions for the reader's convenience. Further details may be found in the original mathematical literature. We then have:
\begin{align}
	\begin{split}
		\mathcal{Z}(q,x)& = \frac{1}{m!} 
		\prod_{b=1}^m\oint_{|w_b|=1} \frac{d w_b}{2\pi i\, w_b^{-k+1}} \sum_{\lambda}H_{\lambda}((w/x)^{-1}\,, q)\Delta(w) \mathcal{Z}_{W}(w) \frac{\langle P_{\lambda},\, P_{k^m}\rangle_P}{(1-q)^m} \\
		& = \frac{1}{m! \varphi_m(q)} \prod_{b=1}^m
		\oint_{|w_b|=1} \frac{d w_b}{2\pi i\, w_b^{-k+1}} H_{k^m}((w/x)^{-1}\,, q)\Delta(w) \mathcal{Z}_{W}(w) \\
		& =  \frac{1}{m! \varphi_m(q)} 
		\sum_{\lambda}\prod_{b=1}^m\oint_{|w_b|=1} \frac{d w_b}{2\pi i\, w_b^{-k+1}} K_{  \lambda,\, k^m}( q)s_{\lambda} ((w/x)^{-1})\Delta(w) \mathcal{Z}_{W}(w) \,,
	\end{split}
\end{align}
where we have used the orthogonality relation among the Hall-Littlewood polynomials \eqref{eq:appPP}. Using the expansion of the Schur polynomials $s_{\lambda}(y)$ in terms of Kostka polynomials we now obtain:
\begin{align}
	\begin{split}
		\mathcal{Z}(q,x)& = \frac{1}{m! \varphi_m(q)} 
		\prod_{b=1}^m\oint_{|w_b|=1} \frac{d w_b}{2\pi i\, w_b^{-k+1}} \Delta(w) \mathcal{Z}_{W}(w)\sum_{\lambda, \nu}  K_{ \lambda,\,  k^m}( q) K_{\lambda, \, \nu}(q)P_{\nu}((w/x)^{-1}, \, q)\, .
	\end{split}
\end{align}
Using \eqref{eq:wkP} and the homogeneity property $P_{\nu}(\alpha\zeta,\, q)= \alpha^{|\nu|}P_{\nu}(\zeta, \, q)$ 
we have:
\begin{align}
	\begin{split}
		\mathcal{Z}(q,x)& = \frac{1}{m! \varphi_m(q)} 
		\prod_{b=1}^m\oint_{|w_b|=1} \frac{d w_b}{2\pi i\, w_b} \Delta(w) \mathcal{Z}_{W}(w)\sum_{\lambda, \nu}  K_{\lambda, \,k^m}( q) K_{\lambda, \, \nu}(q)P_{\nu}((w/x)^{-1}, \, q) P_{k^m}(w, \, q)\, \\
		& = \frac{1}{m! \varphi_m(q)} 
		\prod_{b=1}^m\oint_{|w_b|=1} \frac{d w_b}{2\pi i\, w_b} \Delta(w) \mathcal{Z}_{W}(w)\sum_{\lambda, \nu}  K_{\lambda, \, k^m}( q) K_{\lambda, \, \nu}(q)x^{ |\nu| }P_{\nu}(w^{-1}, \, q) P_{k^m}(w, \, q) \, ,
	\end{split}
\end{align}
then
\begin{align}
	\begin{split}
		\mathcal{Z}(q,\, x) & = \frac{1}{\varphi_m(q)} 
		\sum_{\lambda, \nu}  K_{\lambda,\,k^m}( q) K_{\lambda, \, \nu}(q)x^{|\nu|}\frac{\langle P_{\nu}, \, P_{k^m}\rangle_q}{(1-q)^m} \\
		& = \frac{x^{ k m}}{\varphi_m(q)^2} 
		\sum_{\lambda}  K_{\lambda, \, k^m}(q)^2\,.
	\end{split}
\end{align}
Since there is only one partition $|\lambda| = k m$ and $\ell(\lambda) \leq 1$  we have:
\begin{align} \label{eq:ZT11}
	\mathcal{Z}(q,x)& =   \frac{x^{ km}}{\varphi_m(q)^2} 
	K_{k^m, \,k m}( q)^2 = x^{ k m} q^{k m (m-1)} \prod_{j=1}^m \frac{1}{(1- q^j)^2}\,.
\end{align}


Using the result of section \ref{sec:Bifundamental}, specifically \eqref{eq:ZT11} for $k=1$, and adding the contribution from the vacuum energy $\mathcal{E}_0=\frac{3}{4}N +\frac{3}{4}N= \frac{3}{2}N$, we obtain:
\begin{align}
    \mathcal{Z}(q, \, x) & =  x^{ m} q^{m N}
    q^{ m (m-1)} \prod_{j=1}^m \frac{1}{(1- q^j)^2} \,.
\end{align}
As in the case of $S^5$, an index can be obtained if we shift the chemical potential by $\frac{i \pi}{\gamma }$, hence $x^j \rightarrow {\rm e}^{\gamma\mu  (j+ \frac{i \pi}{\gamma \mu}j)}=(-1)^j x^{j}$. We define:
\begin{align}\label{eq:ImT11x}
\begin{split}
    \mathcal{I}_{U(m)_1\times U(m)_{-1}}(q, x) & = q^{m N}\text{Tr} (-1)^j q^{n}x^{j} \\
    & =  (-1)^mx^{ m} q^{m N}q^{ m (m-1)} \prod_{j=1}^m \frac{1}{(1- q^j)^2} \,.
    \end{split}
\end{align}
We impose again \eqref{eq:R23} which implies
$x \rightarrow {\rm e}^{- \gamma \frac{3}{2}} =q$. Therefore we obtain
\begin{align}
	\begin{split}
		\mathcal{I}_{U(m)_1\times U(m)_{-1}}(q)& = (-1)^m q^{m N}q^{ m^2} \prod_{j=1}^m \frac{1}{(1- q^j)^2}\,.
       	\end{split}
\end{align}
We notice that this index satisfies a factorization property like the index studied in \cite{Fujiwara:2023azx}, however the details are different as we are only probing a specific sector within the zero baryonic charge spectrum and our index is blind to the other global charges. It would be interesting to explore if it would be possible to reproduce fully the results of \cite{Fujiwara:2023azx} by a more systematic analysis of fluctuations along the lines of \cite{Deddo:2025lfm}.



\section{Conclusions}\label{Sec:Conclusion}

In this work, we have revisited the holographic giant graviton expansion from the perspective of the intrinsic gravitational dynamics of D3-branes probing AdS$_5 \times$SE$_5$. Our main goal has been to isolate universal features of the semiclassical quantization of supersymmetric giant gravitons and to clarify their relation to protected indices through the giant graviton expansion.

A central outcome of our analysis is the identification of a universal local mechanism governing fluctuations around supersymmetric embeddings. By reformulating the $S^5$ geometry in terms of K\"ahler and toric data, we showed that the dynamics of fluctuations around maximal giant graviton configurations reduces to an effective Fock-Darwin system. In the appropriate regime, this further collapses to a Landau-level problem whose lowest Landau level encodes the relevant supersymmetric degrees of freedom. The resulting Hilbert space carries a natural representation of the angular momentum operator, which remains robust under deformations away from the strictly supersymmetric point.

This structure provides a unified description of the protected giant graviton sector in the case of AdS$_5 \times S^5$, where the entire giant graviton contribution is captured by a single toric fixed point. More generally, for SE$_5$ geometries such as $T^{1,1}$, the presence of multiple toric corners suggests a decomposition into a collection of analogous local quantum mechanical systems, each associated with a distinct fixed point of the toric action. In this sense, the giant graviton-like construction provides a universal ``patchwise'' quantization of protected BPS sectors in toric AdS$_5\times$SE$_5$ backgrounds.

Importantly, our analysis does not attempt to reconstruct the full global BPS Hilbert space in general SE$_5$ backgrounds. Such a reconstruction would require a consistent prescription for gluing local contributions associated with different toric patches. Instead, we have isolated the fundamental local building blocks that such a construction must be based on.

This viewpoint naturally connects with the interpretation of giant gravitons as holomorphic probes of the Calabi-Yau geometry, as originally advocated in \cite{Mikhailov:2000ya,Beasley:2002xv}. It also aligns with more recent developments in equivariant localization \cite{BenettiGenolini:2026hmz}, where supersymmetric observables like on shell action of the brane are reconstructed from local geometric data.

This perspective also naturally resonates with the open-closed-open triality structure recently emphasized in \cite{Gopakumar:2022djw,Gopakumar:2024jfq}, where gravitational dynamics, open string degrees of freedom, and dual field-theoretic quantities are intertwined. In the present setting, giant gravitons provide a concrete realization of the open-string sector, while the local Landau/Fock-Darwin systems capture its effective semiclassical organization. From this viewpoint, our construction can be understood as a purely gravitational implementation of one leg of this triality, formulated directly in terms of semiclassical D3-brane dynamics.

Several extensions of this framework are natural. Extending the present analysis to more general supersymmetric backgrounds, including AdS$_4 \times$SE$_7$, may reveal further universality in the structure of the emergent Landau-type systems. More broadly, the connection between giant gravitons, bubbling geometries, and black hole microstate counting suggests that the local quantization mechanism identified here may provide a useful building block for a microscopic understanding of supersymmetric gravitational entropy.

\section*{Acknowledgments}
We thank Davide Cassani, Diego Correa, Ignacio Cruz, Evan Deddo,  Alberto Faraggi, Sabare Jayaprakash, Seok Kim, Hyojoong Kim, Yixuan Li, Jim Liu,Sameer Murthy, Mart\'i  Rossell\'o,  Diego Trancanelli. The work of AGL is supported by the University of Padua,, programme ``Quantum gravity, Holography and Black Holes'' with funds of the project PRIN 2022 cod. MUR 2022YZ5BA2 `` Effective quantum gravity'' CUP code C53D23001590006 (co-financed by the project PRIN2017, CUP code C53D23001590006).
LAPZ is partially supported by the U.S.
Department of Energy under grant DE-SC0007859. AR is supported by the National Research Foundation of Korea grant 2021R1A2C2012350. 

\appendix

\section{The Quantum Hall effect partition function}\label{App:QHE}

It will be convenient to reproduce here the procedure leading to the partition function of the type of matrix model associated with the Quantum Hall effect studied in \cite{Dorey:2016hoj}. To this end, we consider a more generic situation where the number of $\Phi_i$ degrees of freedom is $p\geq 1$ and they transform in the fundamental representation of the global $SU(p)$ symmetry group. Then the full partition function is given by
\begin{align}
    \mathcal{Z}(q, \vec{x})& = \text{Tr} \,q^n \prod_{i=1}^p x_i^j, \quad \textrm{with} \quad q \equiv {\rm e}^{- \gamma \Omega^2}, \quad x_i \equiv {\rm e }^{\gamma \mu_i},  \quad \vec{x} \equiv \{x_1, \cdots, x_p\} \,.
\end{align}

To construct the full Hilbert space, we first ignore the $U(m)$ gauge constraints
and introduce fugacities $z_a$ for the Cartan $U(1)^m \subset U(m)$.
Adjoint operators $Z^\dagger_{ab}$ carry weight $z_a/z_b$ and
contribute
\begin{align}
\mathcal{Z}_Z = \prod_{a,b=1}^m \frac{1}{1 - q\, z_a/z_b}\,.
\end{align}
Fundamental fields $\Phi^\dagger_{a i}$ carry weight $z_a x_i$ and contribute
\begin{align}
\mathcal{Z}_\Phi = \prod_{a=1}^m \prod_{i=1}^p \frac{1}{1 - z_a x_i}\,.
\end{align}

Physical states must be invariant under $SU(m)$ and carry fixed charge $k$ under
the $U(1)$ center.  Projecting onto these states by contour integration yields
\begin{align} \label{eq:Zapp}
\mathcal{Z}(q,x_i) = \frac{1}{m!}
\prod_{a=1}^m \oint_{|\,z_a|=1} 
\frac{dz_a}{2\pi i\, z_a^{\,k+1}}
\;\prod_{b\neq c}\!\left(1 - \frac{z_b}{z_c}\right)
\; \mathcal{Z}_Z\, \mathcal{Z}_\Phi \,,
\end{align}
where the product over $b\neq c$ is the Haar measure of $U(m)$ given in terms of  the  Vandermonde determinant $\Delta(z)=\prod_{b\neq c}\!\left(1 - \frac{z_b}{z_c}\right)$.

Because the integrand is a symmetric function in both the $\{z_a\}$ and $\{x_i\}$, the
evaluation proceeds by expanding in a basis of symmetric polynomials. Let $\lambda$ be a partition of a natural number $n\in \mathbb{N}$ such that 
\begin{eqnarray}
     |\lambda| = \lambda_1 + \lambda_2 + ... + \lambda_k = n\,,
\end{eqnarray} where $\{\lambda_i\}$ forms a non-increasing sequence and length $\ell(\lambda) = k$. 
Using standard properties of the space of symmetric polynomials, in \cite{Dorey:2016hoj} the authors evaluated the partition function \eqref{eq:Zapp}. We will follow closely section 3 of \cite{Dorey:2016hoj} to derive the partition function. This will be useful for us to define notation and introduce several special functions that we use in the main body. 
A symmetric function $f(\zeta)=f(\zeta_1, \cdots, \zeta_m)$ is a polynomial in $\{\zeta_i\}\}$ that is invariant under the action of the permutation group $S_m$ on $\{\zeta_i\}$, that is:
\begin{align}
	\sigma \left\{f(\zeta)\right\} \equiv f(\zeta_{\sigma(1)}, \, \cdots, \, \zeta_{\sigma(m)}) = f(\zeta), \quad \forall \, \sigma \in S_m \,.
\end{align}
The Schur functions furnish a basis for the space of symmetric functions and they are defined for every partition $\lambda$ as:
\begin{align}
	s_{\lambda}(\zeta) =\sum_{\sigma \in S_m}\sigma \left\{\zeta_1^{\lambda_1}\cdots \zeta_{m}^{\lambda_m} \prod_{i>j}\frac{1}{1- \frac{\zeta_i}{\zeta_j}}\right\}\,.
\end{align}
For any two symmetric functions $f(\zeta), g(\zeta)$ we can define the inner product:
\begin{align}
	\langle f, \, g \rangle_q & \equiv \frac{1}{m!} \left(\prod_{a=1}^m \oint_{|\zeta_a|=1}\frac{d \zeta_a}{2 \pi i \zeta_a}\right)\frac{\prod_{a\neq b}\left(1-  \frac{\zeta_a}{\zeta_b}\right)}{\prod_{a\neq b}\left(1-  q\frac{\zeta_a}{\zeta_b}\right)} f(\zeta)g(\zeta^{-1})\, .
\end{align}
Here $q$ is an arbitrary complex parameter that will be the fugacity for the bulk modes in our matrix model. The set of basis functions that are orthonormal under the inner product $\langle \, , \rangle_q$ are the Hall-Littlewood polynomials defined for each partition $\lambda$ as:
\begin{align}
	P_{\lambda}(\zeta, \, q) & = \frac{1}{\mathcal{N}_{\lambda, q}}\sum_{\sigma \in S_m} \sigma \left\{\zeta_1^{\lambda_1}\cdots \zeta_m^{\lambda_m} \prod_{i>j}\frac{1-q\frac{\zeta_i}{\zeta_j}}{1- \frac{\zeta_i}{\zeta_j}}\right\}\,,
\end{align}
where 
\begin{align}
	\mathcal{N}_{\lambda, q} = \frac{\varphi_{m - \ell(\lambda)} \prod_{j \geq 1} \varphi_{m_j(\lambda)}}{(1-q)^m}\, ,\quad 	\varphi_m =\prod_{j=1}^m (1- q^j) \, ,
\end{align}
 and $m_j(\lambda)$ is the multiplicity of $j\in \mathbb{Z}^{>0}$ in the partition $\lambda$.
The following orthogonality relation holds:
\begin{align} \label{eq:appPP}
	\langle P_{\lambda}, \, P_{\lambda^{\prime}}\rangle_q & = \frac{\delta_{\lambda\, \lambda^{\prime}}}{\mathcal{N}_{\lambda, q}}\, .
\end{align}
Schur functions and Hall-Littlewood polynomials are related via the Kostka Polynomials $K_{\lambda,\, \mu}(q)$ \cite{Kirillov1988TheBA} as follows:
\begin{align} \label{eq:sKP}
	s_{\lambda}(\zeta) &= \sum_{\mu} K_{\lambda, \, \mu}(q) P_{\mu}(\zeta, \, q)\,.
\end{align}
Another important set of functions are the modified Hall-Littlewood polynomials (or the Milne polynomials) $H_{\lambda}(\zeta, q)$ defined by the relation
\begin{eqnarray}
	H_{\lambda} (\zeta,q) = \sum_{\lambda^\prime} s_{\lambda^\prime} (\zeta)K_{\lambda^\prime,\lambda}(q) \,.
\end{eqnarray}
We can now write:
\begin{align}
	\mathcal{Z}_{\Phi} &=\prod_{a=1}^m \prod_{i=1}^p \frac{1}{1 - z_a x_i} = \sum_{\lambda} H_{\lambda}(x, q)P_{\lambda}(z, q)\,,
\end{align}
therefore the partition function \eqref{eq:Zapp} can be evaluated as
\begin{align}
	\mathcal{Z}(q,x_i) =\frac{1}{m!} \sum_{\lambda} H_{\lambda}(x, \, q)
	\prod_{a=1}^m \oint_{|\,z_a|=1} 
	\frac{dz_a}{2\pi i\, z_a}
	\;\Delta(z)
	\; \mathcal{Z}_Z\, \mathcal{Z}_\Phi P_{\lambda}(z, \, q)P_{k^m}(z^{-1}, \, q) \,,
\end{align}
where we have used 
\begin{align} \label{eq:wkP}
	\prod_{a}^m\zeta_a^k & =P_{k^m}(\zeta, \, q) \,.
\end{align}
Here $k^m$ denotes the partition with $m$ non-zero parts each equal to $k$, namely $k m= \sum_{i=1}^m k$. We then have:
\begin{align}
	\mathcal{Z}(q,x_i) = \sum_{\lambda} H_{\lambda}(x, \, q)\frac{1}{(1-q)^m}\langle P_{\lambda}, \, P_{k^m}\rangle_q =\frac{1}{\varphi_m(q)} H_{k^m}(x, \, q)\,.
\end{align}
The final result takes the form:
\begin{align} \label{eq:ZappEvaluated}
    \mathcal{Z}(q, \, x_i) & = \prod_{j=1}^m \frac{1}{1- q^j} \sum_{\lambda} K_{\lambda, k^m}(q) s_{\lambda}(x)\, .
\end{align}
 The properties of these functions imply that the summand will vanish unless $\ell(\lambda) <p$. Notice that, for $p=1, k=1$, \eqref{eq:ZappEvaluated} reduces to
\begin{align} \label{eq:ZQHE}
    \mathcal{Z}(q, x) & = x^{m}q^{\frac{m(m-1)}{2}} \prod_{j=1}^m \frac{1}{1- q^j}\,
\end{align} as expected.


\bibliographystyle{JHEP}
\bibliography{BHLocalization}

\providecommand{\href}[2]{#2}\begingroup\raggedright\begin{thebibliography}{10}

\bibitem{Benvenuti:2004dy}
S.~Benvenuti, S.~Franco, A.~Hanany, D.~Martelli and J.~Sparks, \emph{{An
  Infinite family of superconformal quiver gauge theories with Sasaki-Einstein
  duals}}, \href{https://doi.org/10.1088/1126-6708/2005/06/064}{\emph{JHEP}
  {\bfseries 06} (2005) 064}
  [\href{https://arxiv.org/abs/hep-th/0411264}{{\ttfamily hep-th/0411264}}].

\bibitem{Maldacena:1997re}
J.~M. Maldacena, \emph{{The large N limit of superconformal field theories and
  supergravity}}, {\emph{Adv. Theor. Math. Phys.} {\bfseries 2} (1998) 231}
  [\href{https://arxiv.org/abs/hep-th/9711200}{{\ttfamily hep-th/9711200}}].

\bibitem{Arai:2019aou}
R.~Arai, S.~Fujiwara, Y.~Imamura and T.~Mori, \emph{{Finite $N$ corrections to
  the superconformal index of toric quiver gauge theories}},
  \href{https://doi.org/10.1093/ptep/ptaa023}{\emph{PTEP} {\bfseries 2020}
  (2020) 043B09} [\href{https://arxiv.org/abs/1911.10794}{{\ttfamily
  1911.10794}}].

\bibitem{Arai:2020qaj}
R.~Arai, S.~Fujiwara, Y.~Imamura and T.~Mori, \emph{{Schur index of the ${\cal
  N}=4$ $U(N)$ supersymmetric Yang-Mills theory via the AdS/CFT
  correspondence}},
  \href{https://doi.org/10.1103/PhysRevD.101.086017}{\emph{Phys. Rev. D}
  {\bfseries 101} (2020) 086017}
  [\href{https://arxiv.org/abs/2001.11667}{{\ttfamily 2001.11667}}].

\bibitem{Imamura:2021ytr}
Y.~Imamura, \emph{{Finite-N superconformal index via the AdS/CFT
  correspondence}}, \href{https://doi.org/10.1093/ptep/ptab141}{\emph{PTEP}
  {\bfseries 2021} (2021) 123B05}
  [\href{https://arxiv.org/abs/2108.12090}{{\ttfamily 2108.12090}}].

\bibitem{Gaiotto:2021xce}
D.~Gaiotto and J.~H. Lee, \emph{{The giant graviton expansion}},
  \href{https://doi.org/10.1007/JHEP08(2024)025}{\emph{JHEP} {\bfseries 08}
  (2024) 025} [\href{https://arxiv.org/abs/2109.02545}{{\ttfamily
  2109.02545}}].

\bibitem{Bourdier:2015wda}
J.~Bourdier, N.~Drukker and J.~Felix, \emph{{The exact Schur index of
  $\mathcal{N}=4$ SYM}},
  \href{https://doi.org/10.1007/JHEP11(2015)210}{\emph{JHEP} {\bfseries 11}
  (2015) 210} [\href{https://arxiv.org/abs/1507.08659}{{\ttfamily
  1507.08659}}].

\bibitem{Drukker:2015spa}
N.~Drukker, \emph{{The $ \mathcal{N}=4 $ Schur index with Polyakov loops}},
  \href{https://doi.org/10.1007/JHEP12(2015)012}{\emph{JHEP} {\bfseries 12}
  (2015) 012} [\href{https://arxiv.org/abs/1510.02480}{{\ttfamily
  1510.02480}}].

\bibitem{Choi:2022ovw}
S.~Choi, S.~Kim, E.~Lee and J.~Lee, \emph{{From giant gravitons to black
  holes}}, \href{https://doi.org/10.1007/JHEP11(2023)086}{\emph{JHEP}
  {\bfseries 11} (2023) 086}
  [\href{https://arxiv.org/abs/2207.05172}{{\ttfamily 2207.05172}}].

\bibitem{Kim:2024ucf}
S.~Kim and E.~Lee, \emph{{Holographic tests for giant graviton expansion}},
  \href{https://doi.org/10.1007/JHEP04(2025)119}{\emph{JHEP} {\bfseries 04}
  (2025) 119} [\href{https://arxiv.org/abs/2402.12924}{{\ttfamily
  2402.12924}}].

\bibitem{Chen:2024erz}
H.-Y. Chen, N.~Dorey, S.~Moriyama, R.~Mouland and C.~Sanli, \emph{{Giant
  gravitons and volume minimisation}},
  \href{https://doi.org/10.1007/JHEP08(2025)121}{\emph{JHEP} {\bfseries 08}
  (2025) 121} [\href{https://arxiv.org/abs/2412.05357}{{\ttfamily
  2412.05357}}].

\bibitem{Lee:2023iil}
J.~H. Lee, \emph{{Trace relations and open string vacua}},
  \href{https://doi.org/10.1007/JHEP02(2024)224}{\emph{JHEP} {\bfseries 02}
  (2024) 224} [\href{https://arxiv.org/abs/2312.00242}{{\ttfamily
  2312.00242}}].

\bibitem{Eleftheriou:2023jxr}
G.~Eleftheriou, S.~Murthy and M.~Rossell{\'o}, \emph{{The giant graviton
  expansion in $AdS_5 \times S^5$}},
  \href{https://doi.org/10.21468/SciPostPhys.17.4.098}{\emph{SciPost Phys.}
  {\bfseries 17} (2024) 098}
  [\href{https://arxiv.org/abs/2312.14921}{{\ttfamily 2312.14921}}].

\bibitem{Beccaria:2024vfx}
M.~Beccaria and A.~Cabo-Bizet, \emph{{Large N Schur index of $ \mathcal{N} $ =
  4 SYM from semiclassical D3 brane}},
  \href{https://doi.org/10.1007/JHEP04(2024)110}{\emph{JHEP} {\bfseries 04}
  (2024) 110} [\href{https://arxiv.org/abs/2402.12172}{{\ttfamily
  2402.12172}}].

\bibitem{Gautason:2024nru}
F.~F. Gautason and J.~van Muiden, \emph{{One-loop quantization of Euclidean
  D3-branes in holographic backgrounds}},
  \href{https://doi.org/10.1007/JHEP06(2024)073}{\emph{JHEP} {\bfseries 06}
  (2024) 073} [\href{https://arxiv.org/abs/2402.16779}{{\ttfamily
  2402.16779}}].

\bibitem{Lee:2024hef}
J.~H. Lee and D.~Stanford, \emph{{Bulk thimbles dual to trace relations}},
  \href{https://arxiv.org/abs/2412.20769}{{\ttfamily 2412.20769}}.

\bibitem{Eleftheriou:2025lac}
G.~Eleftheriou, S.~Murthy and M.~Rossell{\'o}, \emph{{Localization and
  wall-crossing of giant graviton expansions in AdS$_{5}$}},
  \href{https://doi.org/10.1007/JHEP07(2025)126}{\emph{JHEP} {\bfseries 07}
  (2025) 126} [\href{https://arxiv.org/abs/2501.13910}{{\ttfamily
  2501.13910}}].

\bibitem{Deddo:2024liu}
E.~Deddo, J.~T. Liu, L.~A. Pando~Zayas and R.~J. Saskowski, \emph{{Giant
  Graviton Expansion from Bubbling Geometry: Discreteness from Quantized
  Geometry}}, \href{https://doi.org/10.1103/PhysRevLett.132.261501}{\emph{Phys.
  Rev. Lett.} {\bfseries 132} (2024) 261501}
  [\href{https://arxiv.org/abs/2402.19452}{{\ttfamily 2402.19452}}].

\bibitem{Kinney:2005ej}
J.~Kinney, J.~M. Maldacena, S.~Minwalla and S.~Raju, \emph{{An Index for 4
  dimensional super conformal theories}},
  \href{https://doi.org/10.1007/s00220-007-0258-7}{\emph{Commun. Math. Phys.}
  {\bfseries 275} (2007) 209}
  [\href{https://arxiv.org/abs/hep-th/0510251}{{\ttfamily hep-th/0510251}}].

\bibitem{Gadde:2010en}
A.~Gadde, L.~Rastelli, S.~S. Razamat and W.~Yan, \emph{{On the Superconformal
  Index of N=1 IR Fixed Points: A Holographic Check}},
  \href{https://doi.org/10.1007/JHEP03(2011)041}{\emph{JHEP} {\bfseries 03}
  (2011) 041} [\href{https://arxiv.org/abs/1011.5278}{{\ttfamily 1011.5278}}].

\bibitem{Nakayama:2006ur}
Y.~Nakayama, \emph{{Index for supergravity on AdS(5) x T**1,1 and conifold
  gauge theory}},
  \href{https://doi.org/10.1016/j.nuclphysb.2006.08.012}{\emph{Nucl. Phys. B}
  {\bfseries 755} (2006) 295}
  [\href{https://arxiv.org/abs/hep-th/0602284}{{\ttfamily hep-th/0602284}}].

\bibitem{Eager:2012hx}
R.~Eager, J.~Schmude and Y.~Tachikawa, \emph{{Superconformal Indices,
  Sasaki-Einstein Manifolds, and Cyclic Homologies}},
  \href{https://doi.org/10.4310/ATMP.2014.v18.n1.a3}{\emph{Adv. Theor. Math.
  Phys.} {\bfseries 18} (2014) 129}
  [\href{https://arxiv.org/abs/1207.0573}{{\ttfamily 1207.0573}}].

\bibitem{Mikhailov:2000ya}
A.~Mikhailov, \emph{{Giant gravitons from holomorphic surfaces}},
  \href{https://doi.org/10.1088/1126-6708/2000/11/027}{\emph{JHEP} {\bfseries
  11} (2000) 027} [\href{https://arxiv.org/abs/hep-th/0010206}{{\ttfamily
  hep-th/0010206}}].

\bibitem{Beasley:2002xv}
C.~E. Beasley, \emph{{BPS branes from baryons}},
  \href{https://doi.org/10.1088/1126-6708/2002/11/015}{\emph{JHEP} {\bfseries
  11} (2002) 015} [\href{https://arxiv.org/abs/hep-th/0207125}{{\ttfamily
  hep-th/0207125}}].

\bibitem{McGreevy:2000cw}
J.~McGreevy, L.~Susskind and N.~Toumbas, \emph{{Invasion of the giant gravitons
  from Anti-de Sitter space}},
  \href{https://doi.org/10.1088/1126-6708/2000/06/008}{\emph{JHEP} {\bfseries
  06} (2000) 008} [\href{https://arxiv.org/abs/hep-th/0003075}{{\ttfamily
  hep-th/0003075}}].

\bibitem{Das:2000st}
S.~R. Das, A.~Jevicki and S.~D. Mathur, \emph{{Vibration modes of giant
  gravitons}}, \href{https://doi.org/10.1103/PhysRevD.63.024013}{\emph{Phys.
  Rev. D} {\bfseries 63} (2001) 024013}
  [\href{https://arxiv.org/abs/hep-th/0009019}{{\ttfamily hep-th/0009019}}].

\bibitem{Arapoglu:2003ti}
S.~Arapoglu, N.~S. Deger, A.~Kaya, E.~Sezgin and P.~Sundell, \emph{{Multispin
  giants}}, \href{https://doi.org/10.1103/PhysRevD.69.106006}{\emph{Phys. Rev.
  D} {\bfseries 69} (2004) 106006}
  [\href{https://arxiv.org/abs/hep-th/0312191}{{\ttfamily hep-th/0312191}}].

\bibitem{Ouyang:2002vg}
P.~Ouyang, \emph{{Semiclassical quantization of giant gravitons}},
  \href{https://arxiv.org/abs/hep-th/0212228}{{\ttfamily hep-th/0212228}}.

\bibitem{Deddo:2025lfm}
E.~Deddo, S.~Jayaprakash, J.~T. Liu and L.~A. Pando~Zayas, \emph{{Quantized
  Giant Gravitons as the Periodic Table of Supersymmetric States: D3, M2 and
  M5}},  \href{https://arxiv.org/abs/2509.18252}{{\ttfamily 2509.18252}}.

\bibitem{Biswas:2006tj}
I.~Biswas, D.~Gaiotto, S.~Lahiri and S.~Minwalla, \emph{{Supersymmetric states
  of N=4 Yang-Mills from giant gravitons}},
  \href{https://doi.org/10.1088/1126-6708/2007/12/006}{\emph{JHEP} {\bfseries
  12} (2007) 006} [\href{https://arxiv.org/abs/hep-th/0606087}{{\ttfamily
  hep-th/0606087}}].

\bibitem{Polychronakos:2001mi}
A.~P. Polychronakos, \emph{{Quantum Hall states as matrix Chern-Simons
  theory}}, \href{https://doi.org/10.1088/1126-6708/2001/04/011}{\emph{JHEP}
  {\bfseries 04} (2001) 011}
  [\href{https://arxiv.org/abs/hep-th/0103013}{{\ttfamily hep-th/0103013}}].

\bibitem{Dai:2005hh}
J.~Dai, X.-J. Wang and Y.-S. Wu, \emph{{Dynamics of giant-gravitons in the LLM
  geometry and the fractional quantum Hall effect}},
  \href{https://doi.org/10.1016/j.nuclphysb.2005.09.026}{\emph{Nucl. Phys. B}
  {\bfseries 731} (2005) 285}
  [\href{https://arxiv.org/abs/hep-th/0508177}{{\ttfamily hep-th/0508177}}].

\bibitem{fock1928bemerkung}
V.~Fock, \emph{Bemerkung zur quantelung des harmonischen oszillators im
  magnetfeld}, {\emph{Zeitschrift f{\"u}r Physik} {\bfseries 47} (1928) 446}.

\bibitem{darwin1931diamagnetism}
C.~G. Darwin, \emph{The diamagnetism of the free electron},  in
  \emph{Mathematical Proceedings of the Cambridge Philosophical Society},
  vol.~27, pp.~86--90, Cambridge University Press, 1931.

\bibitem{Drigho-Filho:2017bph}
E.~Drigho-Filho, S.~Kuru, J.~Negro and L.~M. Nieto, \emph{{Superintegrability
  of the Fock-Darwin system}},
  \href{https://arxiv.org/abs/1703.06634}{{\ttfamily 1703.06634}}.

\bibitem{Ivanov:2019rbe}
E.~Ivanov, A.~Nersessian, S.~Sidorov and H.~Shmavonyan, \emph{{Symmetries of
  deformed supersymmetric mechanics on K{\"a}hler manifolds}},
  \href{https://doi.org/10.1103/PhysRevD.101.025003}{\emph{Phys. Rev. D}
  {\bfseries 101} (2020) 025003}
  [\href{https://arxiv.org/abs/1911.06290}{{\ttfamily 1911.06290}}].

\bibitem{Dorey:2016mxm}
N.~Dorey, D.~Tong and C.~Turner, \emph{{Matrix model for non-Abelian quantum
  Hall states}}, \href{https://doi.org/10.1103/PhysRevB.94.085114}{\emph{Phys.
  Rev. B} {\bfseries 94} (2016) 085114}
  [\href{https://arxiv.org/abs/1603.09688}{{\ttfamily 1603.09688}}].

\bibitem{Dorey:2016hoj}
N.~Dorey, D.~Tong and C.~Turner, \emph{{A Matrix Model for WZW}},
  \href{https://doi.org/10.1007/JHEP08(2016)007}{\emph{JHEP} {\bfseries 08}
  (2016) 007} [\href{https://arxiv.org/abs/1604.05711}{{\ttfamily
  1604.05711}}].

\bibitem{Romelsberger:2005eg}
C.~Romelsberger, \emph{{Counting chiral primaries in N = 1, d=4 superconformal
  field theories}},
  \href{https://doi.org/10.1016/j.nuclphysb.2006.03.037}{\emph{Nucl. Phys. B}
  {\bfseries 747} (2006) 329}
  [\href{https://arxiv.org/abs/hep-th/0510060}{{\ttfamily hep-th/0510060}}].

\bibitem{Romelsberger:2007ec}
C.~Romelsberger, \emph{{Calculating the Superconformal Index and Seiberg
  Duality}},  \href{https://arxiv.org/abs/0707.3702}{{\ttfamily 0707.3702}}.

\bibitem{Imamura:2022aua}
Y.~Imamura, \emph{{Analytic continuation for giant gravitons}},
  \href{https://doi.org/10.1093/ptep/ptac127}{\emph{PTEP} {\bfseries 2022}
  (2022) 103B02} [\href{https://arxiv.org/abs/2205.14615}{{\ttfamily
  2205.14615}}].

\bibitem{Balasubramanian:2001nh}
V.~Balasubramanian, M.~Berkooz, A.~Naqvi and M.~J. Strassler, \emph{{Giant
  gravitons in conformal field theory}},
  \href{https://doi.org/10.1088/1126-6708/2002/04/034}{\emph{JHEP} {\bfseries
  04} (2002) 034} [\href{https://arxiv.org/abs/hep-th/0107119}{{\ttfamily
  hep-th/0107119}}].

\bibitem{Gauntlett:2004yd}
J.~P. Gauntlett, D.~Martelli, J.~Sparks and D.~Waldram, \emph{{Sasaki-Einstein
  metrics on S(2) x S(3)}}, {\emph{Adv. Theor. Math. Phys.} {\bfseries 8}
  (2004) 711} [\href{https://arxiv.org/abs/hep-th/0403002}{{\ttfamily
  hep-th/0403002}}].

\bibitem{Martelli:2004wu}
D.~Martelli and J.~Sparks, \emph{{Toric geometry, Sasaki-Einstein manifolds and
  a new infinite class of AdS/CFT duals}},
  \href{https://doi.org/10.1007/s00220-005-1425-3}{\emph{Commun. Math. Phys.}
  {\bfseries 262} (2006) 51}
  [\href{https://arxiv.org/abs/hep-th/0411238}{{\ttfamily hep-th/0411238}}].

\bibitem{Forcella:2008bb}
D.~Forcella, A.~Hanany, Y.-H. He and A.~Zaffaroni, \emph{{The Master Space of
  N=1 Gauge Theories}},
  \href{https://doi.org/10.1088/1126-6708/2008/08/012}{\emph{JHEP} {\bfseries
  08} (2008) 012} [\href{https://arxiv.org/abs/0801.1585}{{\ttfamily
  0801.1585}}].

\bibitem{Arean:2004mm}
D.~Arean, D.~E. Crooks and A.~V. Ramallo, \emph{{Supersymmetric probes on the
  conifold}}, \href{https://doi.org/10.1088/1126-6708/2004/11/035}{\emph{JHEP}
  {\bfseries 11} (2004) 035}
  [\href{https://arxiv.org/abs/hep-th/0408210}{{\ttfamily hep-th/0408210}}].

\bibitem{Canoura2006}
F.~Canoura, J.~D. Edelstein, L.~A. Pando~Zayas, A.~V. Ramallo and D.~Vaman,
  \emph{{Supersymmetric branes on $AdS(5) x\times Y^{(p,q)}$ and their field
  theory duals}}, {\emph{JHEP} {\bfseries 03} (2006) 101}
  [\href{https://arxiv.org/abs/hep-th/0512087}{{\ttfamily hep-th/0512087}}].

\bibitem{Canoura:2006es}
F.~Canoura, J.~D. Edelstein and A.~V. Ramallo, \emph{{D-brane probes on
  L(a,b,c) superconformal field theories}}, {\emph{JHEP} {\bfseries 09} (2006)
  038} [\href{https://arxiv.org/abs/hep-th/0605260}{{\ttfamily
  hep-th/0605260}}].

\bibitem{Berenstein:2002ke}
D.~Berenstein, C.~P. Herzog and I.~R. Klebanov, \emph{{Baryon spectra and AdS
  /CFT correspondence}},
  \href{https://doi.org/10.1088/1126-6708/2002/06/047}{\emph{JHEP} {\bfseries
  06} (2002) 047} [\href{https://arxiv.org/abs/hep-th/0202150}{{\ttfamily
  hep-th/0202150}}].

\bibitem{Hamilton:2010sv}
A.~Hamilton, J.~Murugan and A.~Prinsloo, \emph{{Lessons from giant gravitons on
  $AdS_{5}\times T^{1,1}$}},
  \href{https://doi.org/10.1007/JHEP06(2010)017}{\emph{JHEP} {\bfseries 06}
  (2010) 017} [\href{https://arxiv.org/abs/1001.2306}{{\ttfamily 1001.2306}}].

\bibitem{Fujiwara:2023azx}
S.~Fujiwara, \emph{{Schur-like index of the Klebanov-Witten theory via the
  AdS/CFT correspondence}},  \href{https://arxiv.org/abs/2302.04697}{{\ttfamily
  2302.04697}}.

\bibitem{BenettiGenolini:2026hmz}
P.~Benetti~Genolini, C.~Couzens and A.~L{\"u}scher, \emph{{Probing black holes
  with equivariant localization}},
  \href{https://arxiv.org/abs/2604.26786}{{\ttfamily 2604.26786}}.

\bibitem{Gopakumar:2022djw}
R.~Gopakumar and E.~A. Mazenc, \emph{{Deriving the Simplest Gauge-String
  Duality -- I: Open-Closed-Open Triality}},
  \href{https://arxiv.org/abs/2212.05999}{{\ttfamily 2212.05999}}.

\bibitem{Gopakumar:2024jfq}
R.~Gopakumar, R.~Kaushik, S.~Komatsu, E.~A. Mazenc and D.~Sarkar,
  \emph{{Strings from Feynman Diagrams}},
  \href{https://arxiv.org/abs/2412.13397}{{\ttfamily 2412.13397}}.

\bibitem{Kirillov1988TheBA}
A.~N. Kirillov and N.~Reshetikhin, \emph{The bethe ansatz and the combinatorics
  of young tableaux}, {\emph{Journal of Soviet Mathematics} {\bfseries 41}
  (1988) 925}.

\end{thebibliography}\endgroup
\end{document}